\documentclass[aps,prd,amsmath
,nofootinbib         
,twocolumn         
]{revtex4}
 \usepackage{amsmath}
\usepackage{amsfonts}
\usepackage{amssymb}
\usepackage{dcolumn}
\usepackage{bm}
\usepackage[utf8]{inputenc}
\usepackage[T1]{fontenc}
\usepackage{etoolbox}
\usepackage{mathrsfs}
\usepackage{yfonts}

\allowdisplaybreaks[4]     

\begin{document}

\title{Precision cosmology with exact inhomogeneous solutions of General Relativity: the Szekeres models.}

\author{Marie-No\"{e}lle C\'{e}l\'{e}rier}
\affiliation{Laboratoire Univers et TH\'eories, LUTH, Observatoire de Paris, Universit\'e PSL, Universit\'e Paris Cit\'e, CNRS, 5, Place Jules Janssen, 92190 Meudon, France}
\email{marie-noelle.celerier@obspm.fr}

\date{October 22 2024}

\begin{abstract}
The exact Friedman solution to the field equations of General Relativity (GR) describing a homogeneous and isotropic universe together with its linear and higher-order perturbations are the theoretical roots of the current standard model of cosmology. However, despite its global successes, this standard model currently faces a number of tensions and anomalies, which occur mostly from a mismatch between early and late cosmic time region depictions. Actually, in the era of precision cosmology, the late inhomogeneities in the densities of the Universe components can no more be neglected. Moreover, since GR is fundamentally nonlinear, any linear perturbation theory is doomed to fail at reproducing precisely its features, while higher-order perturbed Friedman- Lema\^itre-Robertson-Walker (FLRW) models cannot either claim the status of exact solution to Einstein's field equations. Fortunately, other GR solutions exist which are best suited for this purpose, e. g., exact inhomogeneous solutions able to represent a matter-cosmological constant dominated universe region. Among these, the Szekeres solution, which is devoid of any symmetry, appears as a proper tool for this purpose. Moreover, since these solutions possess the FLRW model as a homogeneous limit, they can be smoothly matched to the standard representation at the inhomogeneity-homogeneity transition. In this paper, the Szekeres solution and its main interesting properties, as well as the equations needed to use this solution in a cosmological context are presented. Then the prospects for a broader use of its abilities are sketched out. In particular, the use of neural networks is proposed to allow, in the future, the fitting of the huge amount of data becoming available to constrain the arbitrary functions and the constant parameters characterizing the Szekeres model, running for representing our late Universe with an increased precision.
\end{abstract}


\maketitle 

\section{Introduction} \label{intro}

\twocolumngrid

The exact Friedman solution to the field equations of General Relativity (GR) describing a homogeneous and isotropic universe equipped with its linear and higher-order perturbations are the theoretical roots of the current standard model of cosmology. However, besides its global successes and the progresses it has provided in our understanding of the Universe, this standard paradigm faces presently a number of tensions and misunderstandings \cite{A22} which hint at a need for a more accurate theoretical background. 

Two main options are therefore possible. Either one dismisses GR and explores modified gravity theories, while taking care of reproducing all GR successful predictions such as dynamics in the solar system, gravitational waves, lensing, redshift, etc. Or one goes on within the GR framework and considers another exact solution able to satisfy the constraints implied by the full range of cosmological data. This latter option is adopted here.

Since the tensions and anomalies currently encountered seem to occur mostly from a mismatch between early and late cosmic time region depictions, we will consider a cosmological model where the inhomogeneities in the densities of the Universe components are intrinsic. The Szekeres solution which describes a dust dominated spacetime devoid of symmetries appears as a widely general and accurate tool for our purpose. Moreover, this solution exhibits the Friedman-Lema\^itre-Robertson-Walker (FLRW) model as a homogeneous limit. It seems therefore fairly well adapted to represent our Universe, since all the best early time predictions of the standard model can be preserved in its framework.

The Szekeres equations are, of course, more involved than the FLRW ones. Anyhow, the interesting physical properties of the model have lead a number of authors to try to overcome these complications to benefit from such properties. However, up to now, the method used has consisted mainly in assuming {\it a priori} simplified parameters and functions describing a given Szekeres model, designed in order to solve a particular cosmological problem. Then the observable quantities attached to this particular model were calculated and compared to those actually observed. Most of the works were numerically completed.

A few authors considered what they called finding ``the metric of the Universe'' \cite{MC08,H09}. However, at the time such a proposal was done, the numerical calculation power needed to perform this plan was too low and these authors were limited in their legitimate ambition. Even the use of the simpler Lema\^itre-Tolman-Bondi (LTB) models proved unable to reach the goal.

Now, we can benefit from the progresses realized in artificial intelligence and deep learning, to consider again such a program. In the present paper, we display a comprehensive set of observables calculated assuming a Szekeres-type Universe, i. e., using the Szekeres solution to represent the late Universe and its FLRW homogeneous limit at larger scales. Then we show how these observables could be constrained by the cosmological data currently available and by those to be collected in a near future.

The paper is organised as follows. In Sec.\ref{sz}, we present the Szekeres solutions with their main physical properties and, in particular, their null geodesic, redshift, and distance equations. The constraints from currently or soon available cosmological observations are discussed in Sec.\ref{cac}. Some effects observable in inhomogeneous models only are analyzed in Sec.\ref{eoi}. The use of deep learning for processing the data fitting is presented in Sec.\ref{dl}. Section\ref{concl} is devoted to the conclusions. Finally, the Szekeres curvature symbols and tensor components needed for the calculations are displayed in the Appendix.

This paper aims indeed at providing an as comprehensive as possible overview of the cosmological abilities of the Szekeres model, while adopting a novel approach which consists in starting from the observed data and obtaining a constrained universe model. The previously known properties and results thus reminded here are described, discussed and extended to the most general context, with corrections or additions made where needed, see, e. g., Eqs. (\ref{s31}) and (\ref{s79}) or the corrected expressions given in the Appendix. New ideas are also explored, such as in the discussion about perturbation theory given at the end of Sec.\ref{nc} or the warning about the $m$ index of the CMB power spectrum displayed in Sec.\ref{cmb}. Moreover, the cosmographic formalism is adapted to the Szekeres framework in Sec.\ref{dce} and, from the expressions obtained for the multipole components of the series expansion pertaining to this formalism, we are able to show that the Szekeres model can indeed reproduce the main multipole features as seen in different observations.

\section{The Szekeres solutions} \label{sz}

An exact solution to Einstein's field equations for a dust gravitationally sourced spacetime with no symmetry has been found by Szekeres in 1975 \cite{S75a}. Since then, this solution has been (too scarcely) considered in a cosmological context and part of its features has been analyzed. Here, we recall, precise, generalize, complete, and correct when needed, the main expressions and properties which will be used to build a consistent cosmological work program. Adopting a convenient coordinate system \cite{H96} the line element in synchronous and comoving coordinates, and in geometric units, can be written as
\begin{equation}
\textrm{d}s^2 = - \textrm{d}t^2 + \frac{\left(\Phi_{,r} - \Phi E_{,r}/E\right)^2}{\epsilon - k} \textrm{d}r^2 + \frac{\Phi^2}{E^2}(\textrm{d}p^2 + \textrm{d}q^2), \label{s1}
\end{equation}
where 
\begin{equation}
E(r,p,q) = \frac{S}{2}\left[ \left(\frac{p-P}{S}\right)^2 + \left(\frac{q-Q}{S}\right)^2 + \epsilon \right],  \label{s2}
\end{equation}
with $\Phi$ being a function of $t$ and $r$ and $k$, $S$, $P$, and $Q$ being functions of the $r$ coordinate alone. The parameter $\epsilon$ determines whether the $(p,q)$ 2-surfaces of constant $\{r,t\}$ are unit 2-spheres $(\epsilon= +1)$, unit 2-pseudospheres, i.e., hyperboloids $(\epsilon = -1)$, or 2-planes $(\epsilon=0)$.

The quasispherical Szekeres (QSS) solutions $(\epsilon = +1)$ can be thought of as a set of nonconcentric evolving spheres, which exhibit a dipole distribution in the energy density variation around each sphere. The dipole arises from the $E_{,r}/E$ term. Actually, on each $(p,q)$ 2-sphere, $E_{,r}/E$ vanishes on an ``equator'', is maximal at one ``pole'' and minimal at the other ``pole'', with $E_{,r}/E|_{min} = - E_{,r}/E|_{max}$.

The quasihyperbolic solutions $(\epsilon = -1)$ can be described as a set of evolving right-hyperboloids piled ``nonconcentrically''. In this case also, the hyperboloids display a ``pseudospherical'' dipole whose strength and orientation on each comoving shell is determined by $S(r)$, $P(r)$, and $Q(r)$. However, half of the dipole is located in the hyperboloid sheet which is not free of shell-crossings \cite{H08}. Therefore, if a shell-crossing should occur in a region spanned by the cosmological solution selected by the observational data, this might constitute a severe physical drawback. Hence, after fitting the data, a key verification will be to test whether the achieved model is actually free of shell crossings. The conditions for no shell-crossing have been displayed by Hellaby and Krasi\'nski for spherical foliation ($\epsilon = +1$) \cite{H02} and for pseudospherical ($\epsilon = -1$) and planar ($\epsilon = 0$) foliations \cite{H08}, and they will have to be checked.

The quasiplanar Szekeres solutions $(\epsilon = 0)$ do not have any $E_{,r}/E$ extremum. Therefore, it does not exhibit any dipolelike structure.

Notice that, even though the quasispherical family has been mostly employed up to now in a cosmological context since it is the best understood case, the surfaces $r=\text{const}$ within a space $t=\text{const}$ might exhibit different geometries such as being quasispherical in regions of the space and quasihyperbolic elsewhere, with a zero curvature at the boundaries \cite{H08}. This planar boundary can either be a thin 3-surface or a 2-surface with finite width. If thin, its lack of dipole would not impair the dipole feature of spacetime as a whole, since it would only concern the instantaneous transition from quasispherical to quasihyperbolic dipolelike matter distributions.

More thorough descriptions of the mathematical and physical properties of these solutions can be found in textbooks \cite{P24,B10} and in a bunch of theoretical papers \cite{H02,H08,H96,K08,K12,G17,K11}. We give below a summary of the properties which will be needed for the present study.

The Einstein field equations written with a dust source and a cosmological constant reduce to the following two:
\begin{equation}
\Phi_{,t}^2 = \frac{2 M}{\Phi} - k + \frac{\Lambda}{3} \Phi^2, \label{s3}
\end{equation}
\begin{equation}
4 \pi \rho(t,r,p,q) = \frac{M_{,r} - 3 M E_{,r}/E}{\Phi^2 (\Phi_{,r} - \Phi E_{,r}/E)}, \label{s4}
\end{equation}
where $\rho$ is the matter energy density, $M(r)$ is another function of the $r$ coordinate which, in the quasispherical case where $\epsilon = +1$, represents the total active gravitational mass in a shell determined by $r = \text{const}$.

Eq. (\ref{s3}) can be integrated as
\begin{equation}
t - t_B(r) = \int_0^{\Phi} \frac{\text{d}\tilde{\Phi}}{{\sqrt{\frac{2 M}{\tilde{\Phi}} - k + \frac{\Lambda}{3} \tilde{\Phi}^2}}}, \label{s5}
\end{equation}
where $t_B$ is an arbitrary function of $r$ which represents the big bang or crunch time.

A given Szekeres model is therefore {\it a priori} determined by six functions of $r$: $k(r)$, $S(r)$, $P(r)$, $Q(r)$, $M(r)$, and $t_B(r)$, one quasiconstant parameter $\epsilon$ and a cosmological constant $\Lambda$ \footnote{ The term quasiconstant employed here means that $\epsilon$ which can  only have three different possible values can change this value according to the following mechanism: a quasispherical and a quasihyperbolic Szekeres region of spacetime whose constant $r$ sphere radius, constant $r$ hyperboloid radius respectively, diverges at some $r$ value, can be joined at their planar boundaries, provided junctions conditions are verified \cite{H08}. Therefore, even though the values of $\epsilon$ are limited to a set of three numbers, its actually realized value depends on that of the $r$ coordinate. The parameter $\epsilon$ can therefore be viewed as a step function of $r$}. However, the number of independent functions of $r$ can be reduced to five by using the coordinate freedom to rescale $r$, e. g., through a choice of any one among the arbitrary functions. 

To construct a Szekeres model able to reproduce a sufficiently large set of cosmological data, we need to express the observable quantities with the tools available to describe the Szekeres solutions. This is the main purpose of the present paper.

\subsection{Null geodesic and redshift equations} \label{null}

The data obtained by cosmological surveys currently available and those which will be displayed in a near future are and will be mainly obtained by measuring electromagnetic signals from astrophysical objects located at some distance from the observer \footnote{Of course, gravitational waves are intended to become a key tool in cosmology, but their consideration is premature and therefore out of the scope of the present study.}. To process them in view of constraining a cosmological model, the trajectories of the photons emitted by the source and measured by the observer in this model, as well as a method for calculating the redshift, need to be known. We give below the main equations describing these features in the most general Szekeres model. The expressions for the geometrical quantities used for establishing these equations are given in the Appendix.

In the geometric optics approximation, based on the assumption that light wave-lengths are negligible with respect to the space curvature scales, which is a reasonable assumption in a cosmological framework, photons travel along null geodesics, and the light rays have no effect on the geometry, i.e., they are test objects. The tangent vector to the null geodesics, $k^{\alpha}$, is given by the gradient of the wave phase and is identified with the photon momentum. It satisfies $k^{\alpha} k^{\beta}_{;\beta} = 0$ and $k^{\alpha} k_{\alpha}=0$. The first equation is the null geodesic equation which can be written, in an affine parametrization as
\begin{equation}
\frac{\text{d}^2 x^{\alpha}}{\text{d}s^2} + \Gamma^{\alpha}_{\beta\gamma} \frac{\text{d}x^{\beta}}{\text{d}s} \frac{\text{d}x^{\gamma}}{\text{d}s} = 0. \label{s6}
\end{equation}
Inserting into (\ref{s6}) the non-zero values of the Christoffel symbols displayed in the Appendix, we obtain the four null geodesic equations as \cite{K11,N11}

\onecolumngrid

\begin{equation}
\frac{\text{d}^2 t}{\text{d}s^2} + \left( \frac{\Phi_{,tr} - \Phi_{,t} E_{,r}/E}{\epsilon - k}\right)\left(\Phi_{,r} - \Phi E_{,r}/E \right)\left(\frac{\text{d}r}{\text{d}s}\right)^2 + \frac{\Phi \Phi_{,t}}{E^2}\left[\left(\frac{\text{d}p}{\text{d}s}\right)^2 + \left(\frac{\text{d}q}{\text{d}s}\right)^2\right] = 0,  \label{s7}
\end{equation}
\begin{eqnarray}
\frac{\text{d}^2r}{\textrm{d}s^2} &+& 2\left(\frac{\Phi_{,tr} - \Phi_{,t} E_{,r}/E}{\Phi_{,r} - \Phi E_{,r}/E}\right)\frac{\text{d}t}{\text{d}s}\frac{\text{d}r}{\text{d}s} + \left(\frac{\Phi_{,rr} - \Phi E_{,rr}/E}{\Phi_{,r} - \Phi E_{,r}/E} - \frac{E_{,r}}{E} + \frac{k_{,r}}{2(\epsilon-k)}\right)\left(\frac{\text{d}r}{\text{d}s}\right)^2 \nonumber \\
&+& 2 \frac{\Phi}{E^2}\left(\frac{E_{,r}E_{,p} - E E_{,r p}}{\Phi_{,r} - \Phi E_{,r}/E}\right) \frac{\text{d}r}{\text{d}s}\frac{\text{d}p}{\text{d}s} + 2 \frac{\Phi}{E^2}\left(\frac{E_{,r}E_{,q} - E E_{,r q}}{\Phi_{,r} - \Phi E_{,r}/E}\right) \frac{\text{d}r}{\text{d}s}\frac{\text{d}q}{\text{d}s} \nonumber \\
&-&  \frac{\Phi}{E^2}\left(\frac{\epsilon - k}{\Phi_{,r} - \Phi E_{,r}/E}\right)\left[\left(\frac{\text{d}p}{\text{d}s}\right)^2 + \left(\frac{\text{d}q}{\text{d}s}\right)^2\right] = 0, \label{s8}
\end{eqnarray}
\begin{eqnarray}
\frac{\text{d}^2p}{\textrm{d}s^2} &+& 2 \frac{\Phi_{,t}}{\Phi}\frac{\text{d}t}{\text{d}s}\frac{\text{d}p}{\text{d}s} - \frac{\Phi_{,r} - \Phi E_{,r}/E}{\Phi(\epsilon - k)}\left(E_{,r}E_{,p} - E E_{,r p}\right)\left(\frac{\text{d}r}{\text{d}s}\right)^2 \nonumber \\
&+& 2 \frac{\Phi_{,r} - \Phi E_{,r}/E}{\Phi}\frac{\text{d}r}{\text{d}s}\frac{\text{d}p}{\text{d}s} - 2 \frac{E_{,q}}{E}\frac{\text{d}p}{\text{d}s}\frac{\text{d}q}{\text{d}s} + \frac{E_{,p}}{E}\left[- \left(\frac{\text{d}p}{\text{d}s}\right)^2 + \left(\frac{\text{d}q}{\text{d}s}\right)^2\right] = 0, \label{s9}
\end{eqnarray}
\begin{eqnarray}
\frac{\text{d}^2q}{\textrm{d}s^2} &+& 2 \frac{\Phi_{,t}}{\Phi}\frac{\text{d}t}{\text{d}s}\frac{\text{d}q}{\text{d}s} - \frac{\Phi_{,r} - \Phi E_{,r}/E}{\Phi(\epsilon - k)}\left(E_{,r}E_{,q} - E E_{,r q}\right)\left(\frac{\text{d}r}{\text{d}s}\right)^2 \nonumber \\
&+& 2 \frac{\Phi_{,r} - \Phi E_{,r}/E}{\Phi}\frac{\text{d}r}{\text{d}s}\frac{\text{d}q}{\text{d}s} - 2 \frac{E_{,p}}{E}\frac{\text{d}p}{\text{d}s}\frac{\text{d}q}{\text{d}s} + \frac{E_{,q}}{E}\left[\left(\frac{\text{d}p}{\text{d}s}\right)^2 - \left(\frac{\text{d}q}{\text{d}s}\right)^2\right] = 0. \label{s10}
\end{eqnarray}

In addition, we have the first integral of the geodesics, which reads
\begin{equation}
\left(\frac{\text{d}t}{\text{d}s}\right)^2 = \frac{\left(\Phi_{,r} - \Phi E_{,r}/E\right)^2}{\epsilon- k}\left(\frac{\text{d}r}{\text{d}s}\right)^2 + \frac{\Phi^2}{E^2}\left[\left(\frac{\text{d}p}{\text{d}s}\right)^2 + \left(\frac{\text{d}q}{\text{d}s}\right)^2\right]. \label{s11}
\end{equation}

\twocolumngrid

Now, two methods can be used to calculate the redshift. First, we can solve the four second order ordinary differential null geodesic Eqs. (\ref{s7})-(\ref{s10}) for $\{t(s), r(s), p(s), q(s)\}$ where the coefficients are expressed with the metric functions which must be evaluated on the null cones while satisfying the field equations (\ref{s3}) and (\ref{s4}). After integration, we can calculate the tangent vector four components $\{k^t, k^r, k^p,k^q\}$, among which $k^t$ will be used to obtain the redshift $z$, given by
\begin{equation}
 1 + z = \frac{k^t_e}{k^t_o},  \label{s12}
\end{equation}
where the subscripts $e$ and $o$ denote values taken at the emitter and at the observer, respectively. Since we have chosen to use comoving synchronous coordinates, the four-velocity of the matter fluid is $u_{\alpha} = (1,0,0,0)$. The null geodesic tangent vector at the observer can be normalized so that $k^t_o = -1$, which implies
\begin{equation}
 1 + z = - k^t_e.  \label{s13}
\end{equation}

The second method is inspired from that applied to the spherically symmetric Lema\^itre-Tolman-Bondi (LTB) model by H. Bondi \cite{B47} and is adapted here to the Szekeres model. A previous attempt restricted to the case of quasispherical symmetry can be consulted \cite{I08}. We give below the calculation adapted to the most general Szekeres model.

Consider two consecutive wave crests of a light ray directed towards the observer. The first satisfies the null condition (\ref{s11}) at $t(s)$, and the second, at $t(s) + T(s)$, where $T$ is the light ray period. Equation (\ref{s11}) describes the first wave crest motion and the second wave crest obeys
\begin{eqnarray}
\left(\frac{\text{d}(t+T)}{\text{d}s}\right)^2 &=& \frac{\left[\Phi_{,r}(t + T,r) - \Phi(t + T,r)E_{,r}/E\right]^2}{\epsilon- k} \left(\frac{\text{d}r}{\text{d}s}\right)^2 \nonumber \\
&+& \frac{\Phi^2(t + T,r)}{E^2}\left[\left(\frac{\text{d}p}{\text{d}s}\right)^2 + \left(\frac{\text{d}q}{\text{d}s}\right)^2\right]. \label{s14}
\end{eqnarray}

Assuming that the period $T$ of the light wave is negligible with respect to the time over which the light propagates - which is an assumption equivalent to the geometric optics approximation - and following Bondi \cite{B47}, we Taylor expand $\Phi$ and $\Phi_{,r}$ to first order as
\begin{equation}
\Phi(t + T,r) = \Phi(t,r) + \Phi_{,t}(t,r) T, \label{s15}
\end{equation}
\begin{equation}
\Phi_{,r}(t + T,r) = \Phi_{,r}(t,r) + \Phi_{,tr}(t,r) T. \label{s16}
\end{equation}

By substituting (\ref{s15}) and (\ref{s16}) into (\ref{s11}) and (\ref{s14}), then subtracting (\ref{s11}) from (\ref{s14}), we obtain

\onecolumngrid

\begin{equation}
\frac{\text{d}t}{\text{d}s}\frac{\text{d}T}{\text{d}s} = T(s)\left\{\left[\frac{\Phi_{,tr} \Phi_{,r} + \Phi \Phi_{,t} (E_{,r}/E)^2 - (\Phi_{,t} \Phi_{,r} + \Phi \Phi_{,tr})(E_{,r}/E)}{\epsilon -k}\right]\left(\frac{\text{d}r}{\text{d}s}\right)^2 
+ \frac{\Phi \Phi_{,t}}{E^2}\left[\left(\frac{\text{d}p}{\text{d}s}\right)^2 + \left(\frac{\text{d}q}{\text{d}s}\right)^2\right]\right\}. \label{s17}
\end{equation}

Then, we write the redshift as
\begin{equation}
 1 + z(s_e) = \frac{T(s_o)}{T(s_e)}.  \label{s18}
\end{equation}
Taking the derivative of (\ref{s18}) with respect to $s_e$, we get
\begin{equation}
\frac{\text{d}z}{\text{d}s_e} = - \frac{\text{d}T(s_e)}{\text{d}s_e} \frac{1 + z(s_e)}{T(s_e)}. \label{s19}
\end{equation}
Then, we substitute (\ref{s19}) into (\ref{s18}) while dropping the subscripts for better readability and we obtain
\begin{equation}
\frac{\text{d}(\ln(1+z))}{\text{d}s} = - \frac{1}{\frac{\text{d}t}{\text{d}s}}\left\{\left[\frac{\Phi_{,tr} \Phi_{,r} + \Phi \Phi_{,t} (E_{,r}/E)^2 - (\Phi_{,t} \Phi_{,r} + \Phi \Phi_{,tr})(E_{,r}/E)}{\epsilon -k}\right]\left(\frac{\text{d}r}{\text{d}s}\right)^2
+  \frac{\Phi \Phi_{,t}}{E^2}\left[\left(\frac{\text{d}p}{\text{d}s}\right)^2 + \left(\frac{\text{d}q}{\text{d}s}\right)^2\right]\right\}. \label{s20}
\end{equation}

After solving the null geodesic equations, with the two field equations satisfied, we have all the results needed to obtain the redshift from (\ref{s20}).

\subsection{Distance equations} \label{distance}

\twocolumngrid

Consider a congruence of light rays with surface area $\delta S$ emitted by a source $S$ and received by a point observer at $O$. Assume that the $\delta S$ sheet at S is orthogonal to the central ray of the bundle emitted from $S$. This bundle encloses a solid angle $\delta \Omega$ at $O$. The observer area distance $D_A$ \footnote{Not to be confused with the angular diameter distance, defined as the ratio of the linear dimension $l$ of the source to the observed angle $\theta$ subtended at the observer by both light rays encompassing this linear dimension. However, when the distortion effect is small, the area distance can be determined to a sufficient accuracy from the angular diameter distance and we can write $l \simeq D_A \theta$.} is defined by \cite{P24,E71}
\begin{equation}
\delta S = D_A^2 \delta \Omega.
\end{equation}
The image propagation along the light rays is characterized by optical kinematic quantities defined in the observer's screen space, which is the two-dimensional space in the rest frame of the matter four-velocity $u^{\alpha}$ orthogonal to $k^{\alpha}$. Defining the propagation vector $k_{\alpha}$ as $k_{\alpha} = \psi_{,\alpha}$, where $\psi$ is the phase of the light-wave, the light-rays have no vorticity. Furthermore, being geodesics, the light rays experience no acceleration. Only the null expansion and the null shear are non-zero. The null expansion $\theta$, measures the area expansion rate of images. It is given, in terms of the affinely parametrized tangent vector by
\begin{equation}
\theta = k^{\alpha}_{;\alpha}.  \label{s21}
\end{equation}

The rate of change of the observer area distance, henceforth simply named area distance, depends on the expansion rate of the light bundle as
\begin{equation}
\frac{\text{d} (\ln D_A)}{\text{d} \lambda} = \theta = k^{\alpha}_{;\alpha}.  \label{s22}
\end{equation}
Now, we insert into (\ref{s22}) the expression for $k^{\alpha}_{;\alpha}$ obtained with the use of the Christoffel symbols displayed in the Appendix. Then, we apply the total derivative expression to the result and integrate which gives
\begin{eqnarray}
D_A &=& \frac{\Phi^2}{E^2}\left(\frac{\Phi_{,r} - \Phi E_{,r}/E}{\sqrt{\epsilon -k}}\right) \nonumber \\
&\times& \exp{\left[\int^s_{s_o} (k^t_{,t} + k^r_{,r} + k^p_{,p} + k^q_{,q}) \text{d}\lambda \right]}, \label{s23}
\end{eqnarray}
where $s_o$ is the value of the affine parameter at the observer, which can be set to zero.

Thus, the area distance is obtained from (\ref{s23}), where the $k^{\alpha}$ and their derivatives $k^{\alpha}_{,\alpha}$ emerge from (\ref{s7})-(\ref{s11}).

Another means of obtaining an expression for the area distance is to use the Sachs optical equations for the null bundle expansion $\theta$ and its shear $\sigma$ which read \cite{S61}
\begin{equation}
\frac{\text{d}\theta}{\text{d}\lambda}  + \theta^2 +|\sigma|^2 = - \frac{1}{2}R_{\alpha\beta} k^{\alpha} k^{\beta}, \label{s24}
\end{equation}
\begin{equation}
\frac{\text{d}\sigma_{\alpha \beta}}{\text{d}\lambda} + \theta \sigma_{\alpha \beta} = - C_{\alpha\beta\gamma\delta} k^{\gamma} k^{\delta}, \label{s25}
\end{equation}
where $R_{\alpha\beta}$ is the Ricci tensor and $C_{\alpha\beta\gamma\delta}$ is the Weyl tensor.

By using (\ref{s22}) into (\ref{s24}), we obtain 
\begin{equation}
\frac{\text{d}^2 D_A}{\text{d}\lambda^2} = - \left( |\sigma|^2 + \frac{1}{2}R_{\alpha\beta} k^{\alpha} k^{\beta}\right) D_A. \label{s26}
\end{equation}
Thus, the area distance can be obtained from solving (\ref{s26}) together with (\ref{s24}), (\ref{s25}) and the null geodesic equations written for the $k^{\alpha}$s, using the Ricci and Weyl tensor components displayed in the Appendix.

Finally, assuming that the photon number is conserved and that photons travel on unique null geodesics, and since we work in the framework of a metric theory of gravity, we can apply the reciprocity theorem to obtain the value of the luminosity distance of a given source at redshift $z$, $D_L$, as \cite{E33}
\begin{equation}
D_L = (1+z)^2 D_A.  \label{s27}
\end{equation}

\section{Constraints from currently available cosmological observations} \label{cac}

\subsection{Supernova data} \label{dds}

Type Ia supernovae (SNIa) are used in cosmology as standard candles. The apparent bolometric magnitude, $m$, of a standard candle with absolute bolometric magnitude, $M$, at a given redshift, $z$, is a function of $z$ and of the functions and parameters determining the cosmological model. It can be written as \cite{P97}
\begin{equation}
m = \mathscr{M} + 5 \log D_L(z;p_a), \label{s27b}
\end{equation}
where $\mathscr{M}$ is the magnitude zero point, which can be measured from the apparent magnitude and redshift of low-redshift examples of the standard candles, and $p_a$ represents the set of model parameters.

To constrain these parameters with SNIa data, one compares the apparent magnitudes of distant supernovae to local SNIa calibrators whose distance is evaluated through, e. g., measurements of Cepheid variables present in these SN Ia hosts. The set of apparent magnitude and redshift measurements for low-redshift SNIa is used to calibrate (\ref{s27b}), and another set of such measurements at higher redshift is used to determine $p_a$, through (\ref{s23}), (\ref{s26}), and (\ref{s27}).

In the Szekeres case, the parameters of the model are included in the expression for $D_L$ obtained through (\ref{s27}), where the redshift calculation is achieved by applying the method displayed in Sec.\ref{null}.

\subsection{Galaxy number counts} \label{nc}

Here, we adapt to the Szekeres models a reasoning formerly designed for LTB spacetimes \cite{C10}. Denoting $n(z)$ the number density of sources in redshift space per steradian and unit redshift interval, the number of sources observed in a given redshift interval $\text{d}z$ and solid angle $\text{d}\Omega$ is $n \text{d}\Omega \text{d}z$ and the total rest mass betwen $z$ and $\text{d}z$ is
\begin{equation}
\mathcal{M} = 4 \pi m n \text{d}z, \label{s28}
\end{equation}
where $m(z)$ is the average mass per source, assumed to be known from astrophysical considerations.

The total rest mass in a volume element $\text{d}^3 V$, defined as the proper volume on a constant time slice, evaluated on the null cone, can be written as
\begin{equation}
\mathcal{M} = \rho \text{d}^3 V, \label{s29}
\end{equation}
which reads, in a Szekeres framework,
\begin{equation}
\mathcal{M} = 4 \pi\rho \frac{\Phi^2}{E^2}\left(\frac{\Phi_{,r} - \Phi E_{,r}/E}{\sqrt{\epsilon -k}}\right)\text{d}r. \label{s30}
\end{equation}

In the Szekeres solutions, the energy density $\rho$ is given by (\ref{s4}). Inserting the corresponding expression into (\ref{s30}), we obtain
\begin{equation}
4 \pi m n \frac{\text{d}z}{\text{d}r} = \frac{M_{,r} - 3 M E_{,r}/E}{E^2 \sqrt{\epsilon - k}}, \label{s31}
\end{equation}
which gives the average mass density in redshift space $m(z) n(z)$. Since the redshift $z$ can be obtained by the algorithm described in Sec.\ref{null}, the mass density in real space follows and can be compared to the measured data.

Now, using large scale structure observables, the physics of clustering is usually developed in the framework of perturbation theory. One could therefore wonder whether this theory might be extended to the Szekeres domain. First, the behavior of perturbations in a general inhomogeneous spacetime is very complicated. Following the seminal work by Gerlach and Sengupta \cite{G79}, who have designed a gauge-invariant perturbation method for general spherically symmetric spacetimes, other authors have specialized the perturbation framework to particular spherically symmetric inhomogeneous cosmological models, e. g., locally rotationally symmetric spacetimes \cite{C07} or LTB solutions \cite{T97,Z08,C09}. It has been shown by these authors that, compared to those of a FLRW cosmology, perturbations of a LTB background are more involved since they cannot be decomposed into the usual scalar, vector, and tensor modes. Indeed, the reduced symmetry leads these modes to couple. We can therefore presume that this coupling might increase in a Szekeres spacetime whose lack of symmetry is total.

As regards the linear density contrast, defined as
\begin{equation}
\delta \rho = \frac{\rho-\rho_b}{\rho_b}, \label{delrho}  
\end{equation}
where $\rho_b$ is the background density, it is a local non-covariant quantity. A spatially invariant density contrast can be defined as follows \cite{M99,B07}
\begin{equation}
S_{IK} = \int_{\Sigma} \left|\frac{h^{\alpha \beta}}{\rho^I} \frac{\partial \rho}{\partial x^{\alpha}} \frac{\partial \rho}{\partial x^{\beta}}\right| \textrm{d}V 
\end{equation}
where $I \in \mathbb{R}$ and $K \in \mathbb{R} \backslash \{0\}$. This family of density contrast indicators can be considered as local or global depending on the size of $\Sigma$. This quantity not only describes the change of density but also the change of its gradients and of the volume of a perturbed region. So this density indicator describes the evolution of the whole region in a more sophisticated way than the usual $\delta \rho$ of (\ref{delrho}).
 
Now, this reasoning has been developed to be applied to smooth regions of inhomogeneous models. In the present work, we are dealing with the most general Szekeres models where the $\epsilon$ parameter can vary by steps of magnitude unity. This should prevent the use of any perturbation scheme in the sense of perturbation theory. Moreover, the Szekeres solution is already considered here as an exact (in the GR sense) perturbation of a matter dominated FLRW region of the universe, whose defining functions and parameters can be as close as needed to those of its FLRW limit. Hence, the perturbation scheme is neither valid nor useful here.

Indeed, structure growth in Szekeres spacetimes has already be considered in the literature. Illuminating examples have been studied in \cite{B06,B07}. There the cosmological model is a quasi-spherical Szekeres solution and different configurations are studied and evolved along different timescales. Such studies will have to be pursued, but, presumably, as {\it a posteriori} analyses, since one needs to know, even broadly, the features of the cosmological Szekeres model to be able to perform such works.

\subsection{Cosmic microwave background (CMB)} \label{cmb}

\subsubsection{CMB multipoles}

The CMB temperature observed at each
point in the sky can be calculated by generating, at the observer, a null geodesic reaching backward in time the last scattering surface with an initial tangent vector pointing in the direction of the given point in the sky. This is done by integrating the geodesic equations (\ref{s7})-(\ref{s10}). The redshift, which determines the last scattering location on the geodesics (defined by its redshift $z_{ls}$) hence, the CMB temperature seen along any line of sight, will then be obtained through (\ref{s20}) \cite{B13}.

For an accurate calculation of the observed CMB multipoles over the whole sky, geodesics going in every direction will have to be integrated. Then the multipoles will be obtained by computing the coefficients of the spherical harmonic expansion
\begin {equation}
a_{lm} = \int_0^{2\pi} \int_0^{\pi}\frac{\Delta T}{T}Y_{lm}(\theta, \phi) \sin \theta \text{d}\theta\text{d}\phi. \label{s32}
\end{equation}

In particular the CMB dipole will be obtained as
\begin {equation}
D = \sqrt{\sum\limits_{m=-1}^{1}|a_{1m}|^2}. \label{s33}
\end{equation}

Now, in the quasispherical, as well as in the quasihyperbolic Szekeres models, the distribution of mass over each single sphere or quasisphere $\{t= \text{const}, r = \text{const}\}$ has the form of a mass-dipole superposed on a monopole \cite{P24,S85,S75b}. This is not the case of the quasiplanar models, but since, in spacetimes where they might be present, such quasiplanar models can be assumed as infinitly thin, their absence of contribution to a dipole can be ignored. The Szekeres dipole-like contribution to matter density, is therefore given by \cite{B10}
\begin{equation}
 \kappa \Delta \rho = \frac{\chi_{,r} - \chi E_{,r}/E}{\Phi^2(\Phi_{,r}-\Phi E_{,r}/E)}
\frac{6 M \Phi_{,r} - 2 M_{,r} \Phi}{ (\Phi_{,r}\chi-\Phi \chi_{,r})}, \label{s33a}
\end{equation}
where
\begin{equation}
\chi = \frac{1 + P^2 + Q^2}{2 S} + \frac{S}{2}. \label{s33b}
\end{equation}

Moreover, this mass-dipole rotates smoothly from a given $t=\text{const}, r=\text{const}$ quasisphere or quasihyperboloid to neighboring ones. This has been first stressed in \cite{S75b}, and subsequently illustrated in \cite{B13,B20}. Now, there are subcases of the Szekeres solution in which the dipole, while still changing strength from a $t=\text{const}, r=\text{const}$ slice to another, does not rotate. Those are the axisymmetric Szekeres subcases. Given the interesting flexibility of the Szekeres solution, it might constitute a good candidate for analyzing the CMB low multipoles. One could therefore assume that (part of) the dipole measured in the CMB data should be due to the Szekeres geometry of the intervening spacetime region, and use it to constrain the Szekeres parameter functions implied in (\ref{s33a}).

Now, the dipole pertaining to the Szekeres model is an intrinsic dipole. To compare it with the CMB's, we might have to subtract the kinetic dipole due to our motion with respect to the CMB rest frame. However, the kinematic interpretation of the CMB dipole has been questioned by some survey analyses with high statistical confidence \cite{S21}, while other studies have claimed its full kinematic origin \cite{A14}. It must be noted, however, that, in the standard analysis used in the latter, the power spectrum terms $C_l$s are obtained by averaging the $C_{lm}$s over all observed directions around the Planck satellite, since, in an isotropic universe, there is no preferred direction. In an anisotropic universe, conversely, the dependence on the $m$ index must be taken into account. Actually, beware that standard data processings might compromise the use of the CMB dipole as a proper tool for constraining the Szekeres model. However, an unbiased estimator for the kinematic dipole amplitude, proportional to the peculiar velocity of the observer, has been recently proposed \cite{Da23}, and might be used in a near future to estimate the kinetic part of the CMB dipole.

Now, imagine that, in the future, the intrinsic part of the CMB dipole should be properly identified, as well as the intrinsic values of the other low-l multipoles, e. g., from physics at last scattering, then the Szekeres determining functions and parameters would be only constrained from the CMB data through the geometry of the null geodesics followed by the microwave rays between last scattering and the observer. 

However, in the meanwhile, it might become possible to constrain our cosmological model without taking the CMB data into account. Once the proper model is fixed by other data sets, we would be able to use (\ref{s33a}) to deduce the intrinsic dipole and therefore to discriminate between the kinetic and intrinsic dipole, as suggested in Sec.\ref{r}. Finally, the amplitude and direction of the observer's tilt with respect to the Hubble bulk flow would be properly identified.

\subsubsection{Sunyaev Zeldovitch effects}

The Sunyaev Zeldovitch effect (SZE) is the added spectral distortion of the CMB through inverse Compton scattering by high-energy electrons in galaxy clusters, in which the low-energy photons receive an energy boost during their collision with the high-energy cluster electrons \cite{S70}. The SZE can be divided into three different types: the thermal, the kinetic, and the polarization effects.

The thermal Sunyaev Zeldovitch effect (TSZE) refers to the effect arising from the interaction of the CMB photons with electrons which owe their high-energy to their temperature. The TSZE spectral distortion of the CMB expressed as a temperature change $\Delta T_{TSZ}$ at dimensionless frequency $x=h\nu/(K_B T_{CMB})$ is given by
\begin{equation}
\frac{\Delta T_{TSZ}}{T_{CMB}} = f(x) y = f(x) \int n_e \frac{k_B T_e}{m_e c^2} \sigma_T \textrm{d} l, \label{sz1}
\end{equation}
where $T_{CMB}$ is the CMB average temperature, $y$ is the Compton y-parameter, which, for an isothermal cluster, equals the optical depth times the fractional energy gain per scattering, $\sigma_T$ is the Thomson scattering crosssection, $n_e$ is the electron number density, $T_e$ is the electron temperature, $k_B$ is the Boltzmann constant, $m_e c^2$ is the electron rest mass energy, and the integration is along the line-of-sight. As can be seen in (\ref{sz1}), $\Delta T_{TSZ}/T_{CMB}$ is independent of redshift. This unique feature of the TSZE makes it a powerful tool for investigating the high-redshift Universe.

Moreover, it has been shown \cite{B91} that the area distance can be obtained from the normalizations $N_{RJ}$ of the TSZE and $N_X$ of the surface brightness distribution through
\begin{equation}
D_A = \frac{N_{RJ}^2}{N_{X}} \left(\frac{m_e c^2}{k_B T_{eo}}\right)^2 \frac{\Lambda_{eo}}{16 \pi T_{CMB}^2 \sigma_T^2(1 + z)^3}, \label{sz2}
\end{equation}
where $\Lambda_{eo}$ is the X-ray spectral emissivity and $z$ is the redshift of the cluster gas. Equation (\ref{sz2}) can be used to compare the expression of $D_A$ in a Szekeres cosmology, obtained as described in Sec.\ref{distance}, to its measured values as obtained through the right-hand side of (\ref{sz2}).

The main features of the TSZE are: (i) it is a small distortion of the CMB, proportional to the cluster pressure integrated along the line-of-sight, see (\ref{sz1}), (ii) it is redshift-independent, (iii) it possesses a unique spectral signature with a decrease in the CMB intensity at low frequencies and an increase at high frequencies, (iv) the integrated TSZE is proportional to the temperature-weighted mass (the total thermal energy) of the cluster, and mostly interesting for our purpose, (v) through (\ref{sz2}) it allows us to determine the cluster area distance in a cosmology independent way.

The kinetic Sunyaev Zeldovitch effect (KSZE) refers to the Doppler effect on the scattered photons due to the cluster bulk velocity owing to its motion with respect to the CMB rest frame. If a component of the cluster velocity is projected along the line-of-sight from the observer to the cluster, then the Doppler effect yields an observable distortion of the CMB spectrum. It gives a method for calculating the peculiar velocity, valid in the non-relativistic limit. However, we have not found a proper way to use it in a Szekeres framework.

As regards the SZ polarisation effect on the CMB photons, it is predicted as too small for being measurable in a near future.

\subsection{Baryon acoustic oscillations (BAO)} \label{BAO}

Before last scattering, the highly coupled photon-baryon plasma oscillates under the competing effects of gravitational collapse and radiation pressure. The resultant acoustic waves travel in the plasma at the sound speed 
\begin{equation}
c_s(z) = \frac{1}{\sqrt{3\left(1 + \frac{3 \rho_b(z)}{4 \rho_{\gamma}(z)}\right)}}, \label{b1}
\end{equation}
where $\rho_b$ and $\rho_{\gamma}$ are the baryon and radiation energy densities, respectively. At baryon decoupling, i. e., end of the drag epoch, the pressure on the baryons disappears and the baryon acoustic wave is frozen in, while the photons stream freely. An enhanced baryon over-density is thus formed at the distance travelled by the sound wave up to decoupling. This implies a baryon acoustic oscillation (BAO) peak in the galaxy correlation function and wiggles (oscillatory features) in their power spectrum \cite{Hu96,E98}.

This feature being initiated in the early universe, it is not especially subject to initial modifications by the late time inhomogeneities. The only effect such inhomogeneities have on BAO measurements is a bending of the light rays used to establish the galaxy correlation function or power spectrum. The simplest model-independent BAO measurements use the angular 2-point correlation function which involves only the angular separation $\theta$ between galaxy pairs. Using thin enough redshift bins, one measures the angular BAO scale given by
\begin{equation}
\theta_{BAO}(z) = \frac{r_s}{(1+z)D_A(z)}, \label{b2}
\end{equation}
where $r_s$ is the sound horizon at the end of the drag epoch, which can be written as
\begin{equation}
r_s = \int_{z_d}^{\infty} c_s (1+z) \text{d}z. \label{b3}
\end{equation}
Since it is large, around 130 Mpc comoving, the acoustic scale is usually assumed to be conserved along its travel through the late universe inhomogeneities. It is therefore considered as a proper standard ruler. We propose to assume temporarily that this property can be retained in a Szekeres framework.

The outcomes of a typical galaxy survey such as DESI \cite{AD24} are displayed under the form of, either a couple of quantities $\{D_M/r_s, D_H/r_s\}$, where the comoving distances $D_M(z)$ and $D_H(z)$ are defined by $D_M(z) = r_s/\Delta \theta$ and $D_H(z) = r_s/\Delta z$, respectively, or by a single quantity $D_V/r_s$, where the angle-averaged distance $D_V$ is defined by
\begin{equation}
D_V(z) = \left[z D_M^2(z) D_H(z)\right]^{1/3}, \label{b4}
\end{equation}
which is measured in samples with low signal-to-noise, such as quasar surveys. We see that $D_M$ is the comoving distance related to the area distance by the reciprocity theorem which yields: $D_M = (1+z)D_A$.

We have described in Sec.\ref{null} and \ref{distance} methods to compute the redshift and the area distance in a Szekeres framework. The value of $r_s$ is independent of the redshift at which it is measured. Therefore, a set of measurements of the angular BAO scale at different redshifts will yield constraints on the Szekeres functions and parameters from the $(1+z) D_A(z)$ factor in the right-hand side of (\ref{b2}).

Notice, however, that small acoustic scale changes of order some tenth of percent, have been identified from second order effects in perturbation theory and N-body simulations \cite{P08}. These should be taken into account when addressing the most precise stages of cosmological modelling. The primary difficulty proceeds from the anisotropic distortions issuing from the galaxy peculiar velocity, see Sec.\ref{r}. The tackling of this issue in the BAO framework is however beyond the scope of the present paper and is left to future work.

\subsection{Weak lensing} \label{wl}

\subsubsection{Generalities}

In curved spacetimes, Maxwell's equations cannot be solved explicitly, save in cases of high symmetry. In a Szekeres model with no symmetry at all, plane waves do not exist.

However, in the short-wave (WKB) \cite{S92} and gravitational lens \cite{S94} approximations, we can consider a locally plane wave, propagating without interaction with matter, and associated with a hypersurface $S$, orthogonal to a congruence of null geodesics representing light-rays. We spot one fiducial (central) light-ray $\gamma_o$ and consider the geodesic deviation of adjacent light-rays $\gamma$ in the bundle relative to $\gamma_o$.This fiducial ray is described by the null tangent vector field $k^{\mu} = \partial x^{\mu}/ \partial \lambda$, since it is assumed to be parametrized by an affine parameter $\lambda$. For a past-directed light cone, $\lambda = 0$ at the observer and increases while going back in time. We can write, at the observer,
\begin{equation}
(u^{\mu}u_{\mu})_o = -1, \label{s34}
\end{equation}
\begin{equation}
u^{\mu}_o k_{\mu} = 1, \label{s35}
\end{equation}
where $u^{\mu}$ is the four-velocity of the matter fluid.

The light bundle is propagated from the observer along the fiducial ray and the deviation of nearby rays relative to this fiducial ray represent the lensing properties of any intervening mass. Adopting in the following the reasoning displayed by Troxel et al. \cite{T14}, we define a deviation vector field or Jacobi field $Y^{\mu}$ as
\begin{equation}
Y^{\mu}(\vec{\vartheta}, \lambda) = \gamma^{\mu}(\vartheta,\lambda) - \gamma_o^{\mu}(\vartheta=0,\lambda), \label{s36}
\end{equation}
such that $Y$ represents the changing separation between some ray $\gamma$ in the bundle and the fiducial ray $\gamma_o$, from which the relative angular position is $\vartheta$. The four-vector $Y^{\mu}$ is decomposed on a spacelike basis $\{E_1,E_2\}$which spans the plane orthogonal to both $u^{\mu}$ and $k^{\mu}$ in order to complete the tetrad along $\gamma_o$, $u^{\mu}$ resulting from parallely propagating $u^{\mu}_o$ along $\gamma_o$. We can therefore rewrite $Y$ as 
\begin{equation}
Y^{\mu} = \xi_1 E_1^{\mu} +  \xi_2 E_2^{\mu} + \xi_0 k^{\mu}, \label{s37}
\end{equation}
where the screen components $\xi_i(i=1,2)$ change according to the deformation equation
\begin{equation}
\dot{\xi_i} =  S_{ij}\xi_j, \qquad S_{ij} = E_i^{\alpha} k_{\alpha;\beta} E_j^{\beta}, \label{s38}
\end{equation}
where a dot denotes differentiation with respect to the affine parameter $\lambda$. In matrix form, (\ref{s38}) becomes
\begin{equation}
\dot{\mathbf {\xi}} =  \mathcal{S} \mathbf{\xi}. \label{s39}
\end{equation}
The optical deformation matrix $\mathcal{S}$ is composed of the Sach's optical scalars of the beam \cite{S61}, i. e., its expansion rate
\begin{equation}
\theta(\lambda) =  k^{\mu}_{;\mu}, \label{s40}
\end{equation}
and its complex shear rate
\begin{equation}
\sigma(\lambda) =  \frac{1}{2}k_{\mu;\nu}(\lambda) \bar{\epsilon}^{\mu}(\lambda)\bar{\epsilon}^{\nu}(\lambda), \label{s41}
\end{equation}
where $\epsilon^{\mu} = E_1^{\mu} + i E_2^{\mu}$ and an over-bar denotes complex conjugation. Then, the optical deformation matrix becomes
\begin{equation}
\mathcal{S}(\lambda) = \left(
\begin{array}{cc}
   \theta(\lambda) - \mathfrak{R}\sigma(\lambda) & \mathfrak{I}\sigma(\lambda) \\
   \mathfrak{I}\sigma(\lambda) & \theta(\lambda) + \mathfrak{R}\sigma(\lambda)
\end{array} \right). \label{s42}
\end{equation}
Differentiating (\ref{s39}) with respect to $\lambda$ gives
\begin{equation}
\ddot{\xi}(\lambda) = \mathcal{T}(\lambda) \xi(\lambda), \label{s43}
\end{equation}
where the optical matrix $\mathcal{T}$ is given by
\begin{eqnarray}
\mathcal{T} &=&  \mathcal{\dot{S}} + \mathcal{S}^2 \nonumber \\
&=&  \left(
\begin{array}{cc}
   \mathscr{R}(\lambda) - \mathfrak{R}\mathscr{F}(\lambda) & \mathfrak{I}\mathscr{F}(\lambda) \\
   \mathfrak{I}\mathscr{F}(\lambda) & \mathscr{R}(\lambda) + \mathfrak{R}\mathscr{F}(\lambda)
\end{array} \right), \label{s44}
\end{eqnarray}
where, from Sach's transport equations, we have for $\theta$ and $\sigma$ \cite{S61}
\begin{equation}
\dot{\theta} + \theta^2 + |\sigma|^2 = \mathscr{R}, \label{s45}
\end{equation}
\begin{equation}
\dot{\sigma} + 2 \theta\sigma = \mathscr{F}, \label{s46}
\end{equation}
with
\begin{equation}
\mathscr{R} = -\frac{1}{2}R_{\alpha\beta}k^{\alpha}k^{\beta}, \label{s47}
\end{equation}
\begin{equation}
\mathscr{F} = -\frac{1}{2} C_{\alpha \beta\gamma\delta}\bar{\epsilon}^{\alpha}k^{\beta}\bar{\epsilon}^{\gamma}k^{\delta}. \label{s48}
\end{equation}
The source of convergence is written in terms of the Ricci curvature, $\mathscr{R}$, while the source of shear is written in terms of the Weyl curvature, $\mathscr{F}$.

For Szekeres spacetimes, the Ricci and Weyl curvatures can be calculated using the Ricci and Weyl tensor components displayed in the Appendix. 

Equation (\ref{s43}) describes the deformation , or lensing, of an infinitesimal light bundle as it can be seen propagating as $\xi_{i}$ on the screen corresponding to each event parametrized by $\lambda$ along the fiducial ray. Due to the linearity of (\ref{s43}), we can use a Jacobi mapping between $\xi_i$ and the initial angle $\vartheta_i$ between the ray of interest and the fiducial ray, which reads
\begin{equation}
\xi(\lambda) = \mathcal{D}(\lambda) \vartheta. \label{s49}
\end{equation}
The evolution of $\mathcal{D}$ with respect to $\lambda$ obeys
\begin{equation}
\ddot{\mathcal{D}} = \mathcal{T D}, \label{s50}
\end{equation}
where, at $\lambda = 0$, we have $\mathcal{D} = 0$ and $\mathcal{\dot{D}} = 1$. The linearity of (\ref{s43}) implies that its solution $\xi(\lambda)$ is related to its initial value, $\dot{\xi}(0) = \vec{\vartheta}$, by a $\lambda$-dependent linear transformation which can be written as (\ref{s49}). With such a choice of $\lambda$, $\vec{\vartheta}$ is the vectorial angle between $\gamma_o$ and a neighboring ray. The Jacobi map (\ref{s49}) takes infinitesimal changes of ray directions at the observer, back to a screen at an event of $\gamma_o$ determined by the value of $\lambda$. If that event is taken on some source plane, $\mathcal{D}(\lambda)$ corresponds to the properly scaled magnification matrix $\mathcal{A}$ of lens theory. Indeed, since we consider an inhomogeneous model with a limiting background FLRW spacetime, this relationship reads \cite{C12}
\begin{equation}
\mathcal{D} = D^{FLRW}_A \mathcal{A}, \label{s50a}
\end{equation}
where $D^{FLRW}_A$ is the angular diameter distance in the background model and $\mathcal{A}$ can be written as
\begin{eqnarray}
\mathcal{A} =  \left(
\begin{array}{cc}
   1 - \kappa - \gamma_1 & \gamma_2 \\
   \gamma_2 & 1 -\kappa + \gamma_1
\end{array} \right), \label{s50b}
\end{eqnarray}
where the complex-valued lensing shear $\gamma = \gamma_1 + i\gamma_2$ describes the stretching of galaxy images due to lensing, and the convergence $\kappa$ describes a change in size and brightness. It is common to write this in terms of the reduced shear:
\begin{equation}
g_i = \frac{\gamma_i}{1 - \kappa}, \label{s50c}
\end{equation}
which gives
\begin{eqnarray}
\mathcal{A} =  (1 - \kappa)\left(
\begin{array}{cc}
   1 - g_1 & g_2 \\
   g_2 & 1 + g_1
\end{array} \right). \label{s50d}
\end{eqnarray}

Now, denoting a solution to (\ref{s50}) as
\begin{eqnarray}
\mathcal{D} =  \left(
\begin{array}{cc}
   D_{11} & D_{12} \\
   D_{21} & D_{22}
\end{array} \right), \label{s50e}
\end{eqnarray}
we can write the convergence and shear due to the inhomogeneities in the Szekeres region of the Universe as
\begin{equation}
\kappa = 1 - \frac{D_{11} + D_{22}}{2 D_A^{FLRW}}, \label{s50f}
\end{equation}
\begin{equation}
\gamma_1 = \frac{D_{22} - D_{11}}{2 D_A^{FLRW}}, \label{s50g}
\end{equation}
\begin{equation}
\gamma_2 = \frac{D_{12}}{\tilde{D_A}} = \frac{D_{21}}{D_A^{FLRW}}, \label{s50h}
\end{equation}
\begin{equation}
\gamma = \sqrt{\gamma_1^2 + \gamma_2^2}. \label{s50i}
\end{equation}

Comparison of (\ref{s39}) and the derivative of (\ref{s49}) shows that
\begin{equation}
\dot{\mathcal{D}} = \mathcal{S D}, \label{s51}
\end{equation}
which implies that $\mathcal{S}$ can be obtained from $\mathcal{D}$.

Now, from the definition (\ref{s49}) of the Jacobi map, it follows that
\begin{equation}
| \text{det} \mathcal{D}(\lambda)| = \frac{\delta A(\lambda)}{\delta \Omega}. \label{s52}
\end{equation}

This shows that $\text{det} \mathcal{D}(\lambda)$ contains information about the area $\delta A(\lambda)$ as well as about the parity, i.e., the orientation of the beam at $\lambda$ relative to that close to the vertex of the light cone.

\subsubsection{Geodesic deviation in the Szekeres model} \label{gd}

The key components of the optical tidal matrix are the Ricci and Weyl curvatures. Their components are given in the Appendix, after renaming the metric components as \cite{T14}
\begin{equation}
H^2 = \frac{\left(\Phi_{,r} - \Phi E_{,r}/E\right)^2}{\epsilon - k}, \qquad F^2 = \frac{\Phi^2}{E^2}. \label{s53}
\end{equation}
With this renaming, the null geodesic equations become
\begin{equation}
\dot{k}^t + H H_{,t}(k^r)^2 + F F_{,t}\left[(k^p)^2 + (k^q)^2 \right] = 0, \label{s54}
\end{equation}
\begin{eqnarray}
&&H^2 \dot{k}^r + 2 H \dot{H}k^r - H H_{,r}(k^r)^2 \nonumber \\
&-& F F_{,r}\left[(k^p)^2 + (k^q)^2 \right] = 0, \label{s55}
\end{eqnarray}
\begin{eqnarray}
&&F^2 \dot{k}^p + 2 F \dot{F}k^r - H H_{,p}(k^r)^2 \nonumber \\
&-& F F_{,p}\left[(k^p)^2 + (k^q)^2 \right] = 0, \label{s56}
\end{eqnarray}
\begin{eqnarray}
&&F^2 \dot{k}^q + 2 F \dot{F}k^r - H H_{,q}(k^r)^2 \nonumber \\
&-& F F_{,q}\left[(k^p)^2 + (k^q)^2 \right] = 0. \label{s57}
\end{eqnarray}
To which we can add the null vector condition written under the form
\begin{equation}
(k^t)^2 - H^2(k^r)^2 - F^2\left[(k^p)^2 + (k^q)^2 \right] = 0. \label{s58}
\end{equation}
Inserting (\ref{s58}) into (\ref{s54})-(\ref{s57}) and using the definition of the total derivative of $H$ and $F$ with respect to $\lambda$, we obtain
\begin{equation}
\dot{k}^t + \frac{F_{,t}}{F}(k^t)^2 + H^2\left(\frac{H_{,t}}{H} - \frac{F_{,t}}{F}\right)(k^r)^2 = 0, \label{s59}
\end{equation}
\begin{equation}
\dot{k}^r - \left(\frac{H_{,r}}{H} - \frac{F_{,r}}{F}\right)(k^r)^2 + \frac{\dot{H}}{H}k^r - \frac{F_{,r}}{H^2 F}(k^t)^2 = 0, \label{s60}
\end{equation}
\begin{equation}
\dot{k}^p - \frac{H^2}{F^2}\left(\frac{H_{,p}}{H} - \frac{F_{,p}}{F}\right)(k^r)^2 + \frac{\dot{F}}{F}k^r - \frac{F_{,p}}{F^3}(k^t)^2 = 0, \label{s61}
\end{equation}
\begin{equation}
\dot{k}^q - \frac{H^2}{F^2}\left(\frac{H_{,q}}{H} - \frac{F_{,q}}{F}\right)(k^r)^2 + \frac{\dot{F}}{F}k^r - \frac{F_{,q}}{F^3}(k^t)^2 = 0. \label{s62}
\end{equation}
Finally, the screen basis vectors $E_1^{\mu}$ and $E_2^{\mu}$ are defined at the observer in terms of the null vector $k_o^{\mu} = (-1,k_o^r,k_o^p,k_o^q)$, of the comoving velocity of the observer $u_o^{\mu} = (1,0,0,0)$, and of the metric functions in terms of $H$ and $F$. These screen basis vectors must satisfy the orthogonality conditions
\begin{equation}
k_{\mu} E_a^{\mu} = 0, \quad u_{\mu} E_a^{\mu} = 0, \quad E_{a \mu} E_b^{\mu} = 0,\label{s63}
\end{equation}
where $a = \{1,2\}$, and which  imply the initial conditions
\onecolumngrid
\begin{equation}
(E_1^{\mu})_o = \left(0,\frac{F}{H}\sqrt{(k_o^p)^2 + (k_o^q)^2}, -\frac{H}{F}\frac{k_o^rk_o^p}{\sqrt{(k_o^p)^2 + (k_o^q)^2}}, -\frac{H}{F}\frac{k_o^rk_o^q}{\sqrt{(k_o^p)^2 + (k_o^q)^2}}\right),\label{s64}
\end{equation}
\begin{equation}
(E_2^{\mu})_o = \left(0,0, \frac{1}{F}\frac{k_o^p}{\sqrt{(k_o^p)^2 + (k_o^q)^2}}, -\frac{1}{F}\frac{k_o^q}{\sqrt{(k_o^p)^2 + (k_o^q)^2}}\right),\label{s65}
\end{equation}
for non-zero $k_o^p$ and $k_o^q$, which means that the null vectors are not initially radial at the observer. 
\twocolumngrid
Once $E_1^{\mu}$ and $E_2^{\mu}$ are determined at the observer, they are parallely propagated together with $u^{\mu}$ and $k^{\mu}$ along the path of the fiducial light ray. Thus, at any position along this fiducial ray parametrized by $\lambda$, we have similarly the following differential equations to be solved
\begin{equation}
\dot{E_a^{\mu}} + \Gamma^{\mu}_{\alpha\beta} E_a^{\alpha}k^{\beta}= 0, \quad a=\{1,2\}.\label{s66}
\end{equation}
These equations can be written as
\begin{equation}
\dot{E_a^t} + H H_{,t}E_a^rk^r + F F_{,t}\left(E_a^p k^p + E_a^q k^q \right) = 0, \label{s67}
\end{equation}
\begin{eqnarray}
&&\dot{E_a^r} + \frac{\dot{H}}{H}E_a^r + \frac{k^r}{H}\left(H_{,t} E_a^t + H_{,p} E_a^p +  H_{,q} E_a^q\right) \nonumber \\
&-& \frac{F F_{,r}}{H^2}\left(E_a^p k^p + E_a^q k^q\right) = 0, \label{s68}
\end{eqnarray}
\begin{eqnarray}
&&\dot{E_a^p} + \frac{\dot{F}}{F}E_a^p + \frac{k^p}{F}\left(F_{,t} E_a^t + F_{,r} E_a^r +  F_{,q} E_a^q\right) \nonumber \\
&-& \frac{H H_{,p}}{F^2}E_a^r k^r - \frac{F_{,p}}{F} E_a^q k^q = 0, \label{s69}
\end{eqnarray}
\begin{eqnarray}
&&\dot{E_a^q} + \frac{\dot{F}}{F}E_a^q + \frac{k^q}{F}\left(F_{,t} E_a^t + F_{,r} E_a^r +  F_{,p} E_a^p\right) \nonumber \\
&-& \frac{H H_{,q}}{F^2}E_a^r k^r - \frac{F_{,q}}{F} E_a^p k^p = 0. \label{s70}
\end{eqnarray}
In general, (\ref{s50}) has no analytic solution. Therefore, to evaluate the Jacobi matrix $\mathcal{D}$, and, through (\ref{s50a}), the magnification matrix $\mathcal{A}$, we have to solve simultaneously the set of second order differential equations (\ref{s50}) for $\mathcal{D}$ and the sets of first order differential equations, (\ref{s59})-(\ref{s62}) for $k^{\mu}$, and (\ref{s67})-(\ref{s70}) for $E_1^{\mu}$ and $E_2^{\mu}$.

Since weak lensing surveys usually yield, for each observed object: (i) the source galaxy redshift, (ii)the galaxy position on the sky, (iii) reduced shear estimates, the reconstruction of $\mathcal{A}$ can be done from these data, and compared to the theoretical Szekeres lensing quantities.

\subsection{Redshift-space distortions} \label{r}

For observed galaxies, especially for faraway objects, the line-of-sight velocity is coupled with the line-of-sight distance. This phenomenon is referred to as the redshift-space distortion (RSD) effect and must be taken into account for cosmological analyses of galaxy position data.

A galaxy peculiar velocity can be divided into two parts: the peculiar velocity of the galaxy host halo, determined by the large-scale matter distribution, and the galaxy random velocity within the halo. Since the RSD effects at large scale are determined by the large scale density, and, at small scale, by the galaxy virial motion in the halo, one can try to model it separately at these two different scales.

The distance between a galaxy and the observer is measured by its redshift, $z_o$, which includes the cosmological redshift, $z_{cos}$, arising from the Hubble expansion, and the Doppler redshift, $z_{pec}$, caused by the galaxy peculiar velocity along the line-of-sight, $v_{los}$. This can be written as
\begin{equation}
\frac{\lambda_o}{\lambda_e} = 1 + z_o = (1+z_{pec})(1 + z_{cos}). \label{r1}
\end{equation}

On large scales, matter around overdensity regions is falling in and such a stream velocity makes the scatter of $z_o$ for galaxies in this region smaller than that of their $z_{cos}$, resulting in a visual compression effect along the line-of-sight known as the Kaiser effect \cite{K87}. On small scales, some nearby galaxies around the target galaxy can have very large line-of-sight peculiar velocities due to virial motion, and thus will make the scatter of $z_o$ along the line-of-sight for those galaxies much bigger than that of their $z_{cos}$, which makes these galaxies stretch along the line-of-sight. This is known as the Finger-of-God effect \cite{J72,K87}. Both the Kaiser and the Finger-of-God effects are special relativistic. Thus, they cannot be used to constrain a general relativistic model such as Szekeres'. The gravitational redshift distortion, on the other hand, arises from the net gravitational redshift earned by a photon climbing out the potential well of a distant galaxy before falling into that of the Milky Way (or of other structures surrounding the observer) on its way to Earth.

Therefore, accurate estimations of line-of-sight velocities is strongly needed. Eliminating the nonlinear effect of RSD, e. g., in the BAO reconstruction, is indeed important. This implies an accurate determination of the redshift in the framework of the Szekeres model which yields proper equations for the null geodesics and the redshift, $z_{cos}$, see Sec.\ref{null}.

Since a galaxy peculiar velocity is determined by surrounding matter, which is normally traced by neighboring galaxies, one needs to define variables which contain the necessary information to extract the galaxy line-of-sight velocity from survey data, which can be analysed and implemented with the tools of artificial neural networks. This has been done in a recent study completed with a $\Lambda$CDM background \cite{Ch23} which should be adapted to the Szekeres background in future works.

\subsection{Redshift drift} \label{rd}

The redshift drift is the redshift increase or decrease that a comoving observer can measure when looking at the same comoving source at two different instants. This drift proceeds from the fact that the source is measured on the observer's two different past light cones. It occurs in any expanding universe, but its magnitude depends on the geometry of the region travelled by the rays.

This effect has already been analyzed for quasispherical models, first, in the case where the geodesics are axially directed in an axially symmetric Szekeres spacetime \cite{M12}, then, for randomly directed geodesics \cite{M22}. However, in the second analysis, a purely Szekeres effect, ensuring that two rays emitted subsequently by the same source are directed in two different directions, has been missed. This effect has been fully worked out by Krasi\'nski and Bolejko \cite{K11} and their method will be used in Sec.\ref{pd} to study both effects,  the redshift drift and the position drift, i. e., a new effect particular to general light rays travelling in an inhomogeneous spacetime.

Now, the cosmological redshift drift effect has been predicted a long time ago \cite{AS62, MV62}. It has not been much considered since then because of its tiny amplitude, i. e., of the order $10^{-18}$ $s^{-1}$. However, measuring it on a time-scale of a few years is now within reach. Some authors have indeed proposed to use strong-gravitational-lens time delays to complete such a measurement \cite{P17,W21,W22,H23}. This could be done within a very short time, the change in time being supplied by the time delay instead of by waiting for years between observations. We can therefore look at future experiments able to measure it within a reasonable time lapse.

\section{Effects observable in inhomogeneous models only} \label{eoi}

We consider here some physical effects occurring only in inhomogeneous models and exhibiting a particular signature when considered in a Szekeres geometry. Should such effects be observed in our Universe, this might be strong hints in favor of an inhomogeneous cosmology at scales corresponding with those of the effects.

\subsection{Differential cosmic expansion and Hubble flow anisotropies} \label{dce}

\subsubsection{Introduction}

Inhomogeneous cosmological models such as LTB and Szekeres, exhibit differential cosmic expansion, i. e., the inhomogeneous counterpart of the Hubble parameter is no more a function of time only as in FLRW, but it becomes a function of space as well as of time.

In the geometric optics approximation, a formula for the redshift is \cite{E61}
\begin{equation}
\frac{1}{(1+z)^2} \frac{\text{d}z}{\text{d}s} = \frac{\Theta}{3} + \Sigma_{\alpha\beta} n^{\alpha} n^{\beta} + u^{\alpha}_{;\beta} u^{\beta}n_{\alpha}, \label{d1}
\end{equation}
where $u^{\alpha}$ is the matter velocity field, $n_{\alpha}$ is the connecting covector field, locally orthogonal to $u^{\alpha}$, $\Sigma_{\alpha \beta}$ is the shear of the velocity field, and $\Theta = u^{\alpha}_{;\alpha}$ is its expansion. In homogeneous and isotropic models, the shear vanishes and the expansion reduces to $\Theta = 3 H(t)$, H being the Hubble parameter. Once cosmic structures are formed, the expansion field becomes nonuniform, and thus the shear and the acceleration, $\dot{u}^{\alpha} = u^{\alpha}_{;\beta} u^{\beta}$ vanish no more which modifies formula (\ref{d1}) for the redshift.

If the redshift of an observed object can be described solely in terms of a homogeneous expansion and of a Doppler effect, such as
\begin{equation}
(1+z)_o = (1+z)_{FLRW}(1+z)_{Doppler}, \label{d2}
\end{equation}
where the first term on the right-hand side refers to the global homogeneous and isotropic expansion of a background FLRW model and the second term corresponds to a Doppler effect with respect to the FLRW frame which combines the motions of the observer (local boost) and of the observed object (peculiar motion), then the redshift anisotropy is called kinematic. The factor $(1+z)_{FLRW}$ must be defined with respect to the canonical choice of the local Lorentz frame corresponding to a local boost of the local group of galaxies.

If the redshift of an observed source (galaxy-type or CMB) cannot be explained entirely in terms of (\ref{d2}), then nonkinematic effects are present and can be a hint of inhomogeneous geometry.

Distances are also affected by the non vanishing of the acceleration and shear of the velocity field. Now, a modification of the expansion rate implies a modification of the distances. Therefore, a measurement of such effects, if it happens to occur, could be considered as a clear step toward inhomogeneous cosmology. Indeed, generic inhomogeneous cosmological models do not expand such as to keep their spatial curvature, $k$, constant. 

A galaxy survey providing, for each galaxy, its angular position, its luminosity distance, and its redshift can be used to compute the redshift dependence of the anisotropy of the local Universe expansion \cite{B16}. For this purpose, each light ray is propagated, using the null geodesic equations (\ref{s7})-(\ref{s10}), up until $D_L$ equals the measured distance, and the corresponding redshift is obtained through (\ref{d1}). Then the dipole proceeds from the reconstruction of mock catalogues where the measured galaxy redshifts are replaced by their Szekeres counterparts computed in the measured direction. 

The anisotropic expansion is usually encompassed into a decomposition of the quantity corresponding to the FLRW Hubble parameter into a transversal and a longitudinal part,
\begin{equation}
H_{\bot} = \frac{\Phi_{,t}}{\Phi},
\end{equation}
\begin{equation}
H_{\|} = \frac{\Phi_{,tr} - \Phi_{,t}E_{,r}/E}{\Phi_{,r} - \Phi E_{,r}/E},
\end{equation}
respectively. An adaptation of this formalism to the Szekeres case has been made in \cite{V14}. Unfortunately, the results, given in the Goode-Wainwright coordinates \cite{G82}, cannot be used in our framework.

\subsubsection{Cosmography}

Taylor expansions of the luminosity distance in powers of the redshift have long been known under the name ``cosmography''. A simplified version of this formalism has been applied to the homogeneous FLRW model \cite{W72} and to LTB toy models \cite{C00}.

Now, generalized cosmographic expansions of distances have experienced recent progresses, both theoretical \cite{C20,H21,Ka23,M24,A24, K24} and in observation fitting \cite{D23,C23,G24}. From these works, it appears that the most prominent features in an inhomogeneous expansion are the dipolar and quadrupolar moments.

Both quantities have been given a particular description in a general setting in \cite{H21} and we examine below how it can be adapted to a Szekeres universe and what would be the cosmological implications.

It is considered, in \cite{H21}, series expansion of  the luminosity distance $d_L$ in redshift, up to the third order, for general anisotropic spacetimes, provided that they admit such converging series. It reads
\begin{equation}
d_L = d_L^{(1)}z + d_L^{(2)}z^2 + d_L^{(3)}z^3 + \mathcal{O}(z^4), \label{cos1}
\end{equation}
with
\begin{eqnarray}
d_L^{(1)} &=& \frac{1}{\textfrak{H}_o}, \quad d_L^{(2)} = \frac{1}{\textfrak{H}_o}\left(1 + \frac{1}{2 E_o}\frac{\frac{\textbf{d} \textfrak{H}}{\textbf{d} \lambda}}{\textfrak{H}^2} \middle|_{_{_{_{_{_{_{_{_{_{_o}}}}}}}}}}\right) \nonumber \\ 
d_L^{(3)} &=& \frac{1}{\textfrak{H}_o}\left(\frac{1}{2 E_o^2}\frac{\left(\frac{\textbf{d} \textfrak{H}}{\textbf{d} \lambda}\right)^2}{\textfrak{H}^4} \middle|_{_{_{_{_{_{_{_{_{_{_o}}}}}}}}}} - \frac{1}{6 E_o^2}\frac{\frac{\textbf{d}^2 \textfrak{H}}{\textbf{d} \lambda^2}}{\textfrak{H}^3} \middle|_{_{_{_{_{_{_{_{_{_{_o}}}}}}}}}} \right.\nonumber \\
&-& \left. \frac{1}{12 E_o^2}\frac{k^{\mu} k^{\nu} R_{\mu \nu}}{\textfrak{H}^2} \middle|_{_{_{_{_{_{_{_{_{_{_o}}}}}}}}}} \right), \label{cos2}
\end{eqnarray}
where $E_o$ is the photon energy measured at the vertex point $o$ of the observer's light cone, $\lambda$ is an affine parameter along the null geodesics of the photon congruence with tangent vector $k^{\mu}$, satisfying $k^{\mu} \lambda_{,\mu} = 1$, $R_{\mu \nu}$ is the Ricci tensor and
\begin{equation}
\textfrak{H} = \frac{\theta}{3} - e^{\mu} a_{\mu} + e^{\mu} e^{\nu} \sigma_{\mu \nu}, \label{cos3}
\end{equation}
where $\theta$, $a_{\mu}$, and $\sigma_{\mu \nu}$ denote the expansion scalar, the acceleration vector, and the shear tensor of the timelike congruence $u^{\mu}$ of the matter source, respectively. Here, $u^{\mu}$ is chosen to be a unit vector, so that $u^{\mu} u_{\mu} = -1$.

Three other effective observational parameters are then defined as
\begin{equation}
\textfrak{Q} = -1 - \frac{1}{E} \frac{\frac{\textbf{d}\textfrak{H}}{\textbf{d} \lambda}}{\textfrak{H}^2}, \label{cos4}
\end{equation}
\begin{equation}
\textfrak{J} = \frac{1}{E^2} \frac{\frac{\textbf{d}^2\textfrak{H}}{\textbf{d} \lambda^2}}{\textfrak{H}^3} - 4 \textfrak{Q} - 3, \label{cos5}
\end{equation}
\begin{equation}
\textfrak{R} = 1 + \textfrak{Q} -\frac{1}{2 E^2}\frac{k^{\mu} k^{\nu} R_{\mu \nu}}{\textfrak{H}^2}. \label{cos6}
\end{equation}
The expansion coefficients given by (\ref{cos1}) can therefore be written in terms of those parameters evaluated at the vertex point \cite{H21}. To analyze the dependence of these effective kinematical parameters on the observation direction, the luminosity distance Hubble law undergoes a multipole decomposition. The $\textfrak{Q}$ and $\textfrak{J}$ parameters are first modified according to
\begin{equation}
\hat{\textfrak{Q}} = \textfrak{H}^2 \left(\textfrak{Q} + 1 \right), \label{cos7}
\end{equation}
\begin{equation}
\hat{\textfrak{J}} = \textfrak{H}^3 \left(\textfrak{J} - 1 - \textfrak{R} \right). \label{cos8}
\end{equation}
The parameter $\textfrak{H}$ is given by (\ref{cos3}), which can be written as
\begin{equation}
\textfrak{H}(\textrm{\bf e}) = \frac{\theta}{3} - \textrm{\bf e}.\textrm{\bf a} + \textrm{\bf e} \textrm{\bf e} . {\bf \sigma}, \label{cos9}
\end{equation}
where the compact notation $\textrm{\bf e}.\textrm{\bf a} = e^{\mu}a_{\mu}$ and $\textrm{\bf e}\textrm{\bf e} . {\bf \sigma} = e^{\mu}e^{\nu}\sigma_{\mu \nu}$ has been used. The first other effective parameters read
\begin{equation}
\hat{\textfrak{Q}}(\textrm{\bf e}) = - \left(\stackrel{0}{\textfrak{q}}+ \textrm{\bf e}. \stackrel{1}{\textfrak{q}} + \textrm{\bf e} \textrm{\bf e}. \stackrel{2}{\textfrak{q}} + \textrm{\bf e} \textrm{\bf e} \textrm{\bf e}. \stackrel{3}{\textfrak{q}} + \textrm{\bf e} \textrm{\bf e} \textrm{\bf e} \textrm{\bf e}.\stackrel{4}{\textfrak{q}}\right), \label{cos10}
\end{equation}
with
\begin{eqnarray}
\stackrel{0}{\textfrak{q}} &=& \frac{1}{3}\frac{\textrm{d} \theta} {\textrm{d}\tau} + \frac{1}{3} D_{\mu}a^{\mu} - \frac{2}{3}a^{\mu}a_{\mu} -\frac{2}{5} \sigma_{\mu \nu}\sigma^{\mu \nu}, \nonumber \\ 
\stackrel{1}{\textfrak{q}}_{\mu} &=& - h^{\nu}_{\mu}\frac{\textrm{d}a_{\nu}}{\textrm{d}\tau} - \frac{1}{3}D_{\mu}\theta +  a^{\nu}\omega_{\mu \nu}+ \frac{9}{5} a^{\nu}\sigma_{\mu \nu} - \frac{2}{5} D_{\nu} \sigma^{\nu}_{\mu}, \nonumber \\
\stackrel{2}{\textfrak{q}}_{\mu \nu} &=& h^{\alpha}_{\mu}h^{\beta}_{\nu}\frac{\textrm{d}\sigma_{\alpha \beta}}{\textrm{d}\tau} +D_{\langle \mu} a_{\nu \rangle} + a_{\langle \mu} a_{\nu \rangle} - 2 \sigma_{\alpha (\mu}\omega^{\alpha}_{\nu)}  \nonumber \\
&-& \frac{6}{7} \sigma_{\alpha \langle\mu}\sigma^{\alpha}_{\nu \rangle}, \nonumber \\
\stackrel{3}{\textfrak{q}}_{\mu \nu \rho} &=& - D_{\langle \mu} \sigma_ {\nu \rho \rangle} - 3 a_{\langle \mu} \sigma_{\nu \rho \rangle}, \nonumber \\
\stackrel{4}{\textfrak{q}}_{\mu \nu \rho \kappa} &=& 2 \sigma_{\langle \mu \nu} \sigma_{\rho \kappa \rangle}, \label{cos10a}
\end{eqnarray}
where the notation $T_{\langle \mu_1, \mu_2, ..., \mu_n \rangle}$ denotes the symmetric and traceless part of the 3-dimensional tensor \cite{H21}, the operator $\textrm{d}/\textrm{d} \tau = u^{\mu}\nabla_{\mu}$ is the directional derivative along the four-velocity field $u^{\mu}$ of the observer's congruence, and $\theta$, $a^{\mu}$, $\sigma^{\mu \nu}$, and $\omega^{\mu \nu}$ are the hydrodynamical quantities attached to the matter/observer congruence, i. e., the expansion scalar, the acceleration vector, and the shear and vorticity tensors, respectively. The $\hat{\textfrak{J}}(\textrm{\bf e})$ parameter can be developed likewise, but we will not need it in the following.

The above formalism has been used in \cite{M21} to calculate the effective observational Hubble, $\textfrak{H}$, deceleration, $\textfrak{Q}$, jerk, $\textfrak{J}$, and curvature, $\textfrak{R}$ parameters in numerical simulations of realistic structure formation. In \cite{D23,C23}, the same effective parameters have been estimated from different Type Ia supernovae surveys. The results of these analyses imply that the dominant form of anisotropy in the effective Hubble parameter is the quadrupole, i. e., the contribution of the shear tensor dominates the four-acceleration term in (\ref{cos3}), and that the dipole, which is the dominating anisotropic term in the effective deceleration parameter, can be attributed to the first two terms of $\stackrel{1}{\textfrak{q}}_{\mu}$ in (\ref{cos10}), i. e., the spatial gradient of the expansion rate and of the shear tensor.

Such features should therefore be used to constrain the Szekeres model. The kinetic quantities useful in the above equations will thus be considered in a general Szekeres framework. First, we note that the Szekeres acceleration vector and vorticity tensor vanish. The expansion scalar reads
\begin{equation}
\theta = 2 \frac{\Phi_{,t}}{\Phi} + \frac{\Phi_{,tr} - \Phi_{,t}E_{,r}/E}{\Phi_{,r} - \Phi E_{,r}/E}, \label{cos11}
\end{equation}
and the shear tensor is \cite{P24}
\begin{equation}
\sigma^{\mu}_{\nu} = \frac{1}{3} \frac{\Phi_{,tr} - \Phi_{,t}\Phi_{,r}/\Phi}{\Phi_{,r} - \Phi E_{,r}/E} \left(0, 2, -1, -1\right). \label{cos12}
\end{equation}
By inserting (\ref{cos11}), (\ref{cos12}), and $a_{\mu} = 0$ into (\ref{cos3}), we obtain
\begin{eqnarray}
\textfrak{H} &=& \frac{\theta}{3} + \textrm{\bf e} \textrm{\bf e} . {\bf \sigma} = \frac{2}{3} \frac{\Phi_{,t}}{\Phi} + \frac{\Phi_{,tr} - \Phi_{,t}E_{,r}/E}{3\left(\Phi_{,r} - \Phi E_{,r}/E\right)} \nonumber \\
&+& \frac{2}{3}\frac{\left(\Phi_{,tr} - \Phi_{,t}\Phi_{,r}/\Phi\right)\left(\Phi_{,r} - \Phi E_{,r}/E\right)}{\epsilon -k} \left(e^r\right)^2 \nonumber \\
&-& \frac{1}{3} \frac{\Phi^2}{E^2} \left(\frac{\Phi_{,tr} - \Phi_{,t}\Phi_{,r}/\Phi}{\Phi_{,r} - \Phi E_{,r}/E}\right) \left[\left(e^p\right)^2 + \left(e^q\right)^2\right].\label{cos13}
\end{eqnarray}
Similarly we calculate
\begin{eqnarray}
\stackrel{0}{\textfrak{q}} &=& \frac{2}{3} \frac{\textrm{d}}{\textrm{d}\tau} \left(\frac{\Phi_{,t}}{\Phi} \right) + \frac{1}{3} \frac{\textrm{d}}{\textrm{d}\tau} \left(\frac{\Phi_{,tr} - \Phi_{,t}E_{,r}/E}{\Phi_{,r} - \Phi E_{,r}/E} \right) \nonumber \\
&-& \frac{4}{15}\left(\frac{\Phi_{,tr} - \Phi_{,t}\Phi_{,r}/\Phi}{\Phi_{,r} - \Phi E_{,r}/E}\right)^2, \label{cos14}
\end{eqnarray}
\begin{equation}
\stackrel{1}{\textfrak{q}_0} = \frac{2}{3} \frac{\Phi_{,t}^2 - \Phi_{,tt}\Phi}{\Phi^2} - \frac{1}{3} \frac{\partial}{\partial t} \left(\frac{\Phi_{,tr} - \Phi_{,t}E_{,r}/E}{\Phi_{,r} - \Phi E_{,r}/E} \right), \label{cos15}
\end{equation}
\begin{equation}
\stackrel{1}{\textfrak{q}_1} = \frac{2}{3}\left( \frac{\Phi_{,t}\Phi_{,r} - \Phi_{,tr}\Phi}{\Phi^2}\right) - \frac{4}{15} \partial_r \left( \frac{\Phi_{,tr} - \Phi_{,t}\Phi_{,r}/\Phi}{\Phi_{,r} - \Phi E_{,r}/E} \right), \label{cos16}
\end{equation}
\begin{equation}
\stackrel{1}{\textfrak{q}_2} = \frac{1}{15} \frac{\left(E_{,rp}/E - E_{,r}E_{,p}/E^2\right)}{\left(\Phi_{,r} - \Phi E_{,r}/E\right)^2} \left(3 \Phi_{,t} \Phi_{,r} - 2 \Phi \Phi_{,tr} \right), \label{cos17}
\end{equation}
\begin{equation}
\stackrel{1}{\textfrak{q}_3} = \frac{1}{15} \frac{\left(E_{,rq}/E - E_{,r}E_{,q}/E^2\right)}{\left(\Phi_{,r} - \Phi E_{,r}/E\right)^2} \left(3 \Phi_{,t} \Phi_{,r} - 2 \Phi \Phi_{,tr} \right). \label{cos18}
\end{equation}
And as regards the components of $\stackrel{2}{\textfrak{q}}_{\mu \nu}$, they can be written as
\begin{equation}
\stackrel{2}{\textfrak{q}}_{\mu \nu} = \sum\limits_{\stackrel{i \neq 0}{{i\neq \mu},{i\neq \nu}}} \left(u^i u_{\mu}\right)\left(u^i u_{\nu}\right) \frac{\textrm{d}\sigma_{ii}}{\textrm{d}\tau}. \label{cos19}
\end{equation}
We can thus see that a dominant quadrupole in the effective Hubble parameter and a dominant dipole in the effective deceleration parameter as measured in \cite{D23,C23} implies a dominant shear in relation with the volume expansion rate of a Szekeres region. This shows that the cosmographic method is an appropriate and consistent tool for constraining the Szekeres model at low redshifts, since the spacetime properties are consistent in both analyses.

However, the method must be handled with care, since the convergence of the (\ref{cos1}) series must be ensured for every values of the measured redshifts. When the redshift approaches unity or goes beyond, the series convergence radius might be overcome due to the highest power terms. However, we can suspect that the inhomogeneity-homogeneity transition should be reached before. Anyhow, the convergence has to be checked {\it a posteriori}, whatever the redshift range, once the coefficients in the series have been calculated. A means to improve the convergence of this method could be the use of Pad\'e approximants \cite{A24}. However, the choice of the approximants is not unique which adds unwanted new degrees of freedom to the problem.

\subsection{Redshift drift and position drift} \label{pd}

As explained in Sec.\ref{rd} an observer located in a Szekeres universe who measures the same light source at different instants observes not only a shift in the object redshift, but also in its position on the sky. A method for calculating both effects can be inspired from that used by Krasi\'nski and Bolejko \cite{K11}. The aim of these authors was to exhibit the new position drift effect and we will see that, as a bonus, we can obtain the redshift drift.

The null geodesic equations (\ref{s7})-(\ref{s10}) can be written with respect to the $r$ coordinate by implementing the following formula for the total derivative:
\begin{equation}
\frac{\text{d}^2 x^{\alpha}}{\text{d}s^2} = \left(\frac{\text{d}r}{\text{d}s}\right)^2 \frac{\text{d}^2 x^{\alpha}}{\text{d}r^2} + \frac{\text{d}^2 r}{\text{d}s^2}  \frac{\text{d}x^{\alpha}}{\text{d}r}, \label{s80}
\end{equation}
which gives

\onecolumngrid

\begin{equation}
\frac{\text{d}^2 t}{\text{d}r^2} + \left( \frac{\Phi_{,tr} - \Phi_{,t} E_{,r}/E}{\epsilon - k}\right)\left(\Phi_{,r} - \Phi E_{,r}/E \right) + \frac{\Phi \Phi_{,t}}{E^2}\left[\left(\frac{\text{d}p}{\text{d}r}\right)^2 + \left(\frac{\text{d}q}{\text{d}r}\right)^2\right] + U \frac{\text{d}t}{\text{d}r} = 0,  \label{s81}
\end{equation}
\begin{eqnarray}
&&\frac{\text{d}^2p}{\textrm{d}r^2} + 2 \frac{\Phi_{,t}}{\Phi}\frac{\text{d}t}{\text{d}r}\frac{\text{d}p}{\text{d}r} - \frac{\Phi_{,r} - \Phi E_{,r}/E}{\Phi(\epsilon - k)}\left(E_{,r}E_{,p} - E E_{,r p}\right) \nonumber \\
&+& 2 \frac{\Phi_{,r} - \Phi E_{,r}/E}{\Phi}\frac{\text{d}p}{\text{d}r} - 2 \frac{E_{,q}}{E}\frac{\text{d}p}{\text{d}r}\frac{\text{d}q}{\text{d}r} + \frac{E_{,p}}{E}\left[- \left(\frac{\text{d}p}{\text{d}r}\right)^2 + \left(\frac{\text{d}q}{\text{d}r}\right)^2\right] + U \frac{\text{d}p}{\text{d}r}= 0, \label{s82}
\end{eqnarray}
\begin{eqnarray}
&&\frac{\text{d}^2q}{\textrm{d}r^2} + 2 \frac{\Phi_{,t}}{\Phi}\frac{\text{d}t}{\text{d}r}\frac{\text{d}q}{\text{d}r} - \frac{\Phi_{,r} - \Phi E_{,r}/E}{\Phi(\epsilon - k)}\left(E_{,r}E_{,q} - E E_{,r q}\right) \nonumber \\
&+& 2 \frac{\Phi_{,r} - \Phi E_{,r}/E}{\Phi}\frac{\text{d}q}{\text{d}r} - 2 \frac{E_{,p}}{E}\frac{\text{d}p}{\text{d}r}\frac{\text{d}q}{\text{d}r} + \frac{E_{,q}}{E}\left[\left(\frac{\text{d}p}{\text{d}r}\right)^2 - \left(\frac{\text{d}q}{\text{d}r}\right)^2\right] + U \frac{\text{d}q}{\text{d}r}= 0, \label{s83}
\end{eqnarray}
where we have defined
\begin{eqnarray}
U(t,r,p,q) &=& - 2\frac{\Phi_{,tr} - \Phi_{,t} E_{,r}/E}{\Phi_{,r} - \Phi E_{,r}/E}\frac{\text{d}t}{\text{d}r} - \left[\frac{\Phi_{,r r} - \Phi E_{,r r}/E}{\Phi_{,r} - \Phi E_{,r}/E} - \frac{E_{,r}}{E} + \frac{k_{,r}}{2(\epsilon-k)}\right] - 2 \frac{\Phi}{E^2}\left(\frac{E_{,r}E_{,p} - E E_{,r p}}{\Phi_{,r} - \Phi E_{,r}/E}\right) \frac{\text{d}p}{\text{d}r} \nonumber \\
&-& 2 \frac{\Phi}{E^2}\left(\frac{E_{,r}E_{,q} - E E_{,r q}}{\Phi_{,r} - \Phi E_{,r}/E}\right) \frac{\text{d}q}{\text{d}r}
+  \frac{\Phi}{E^2}\left(\frac{\epsilon - k}{\Phi_{,r} - \Phi E_{,r}/E}\right)\left[\left(\frac{\text{d}p}{\text{d}r}\right)^2 + \left(\frac{\text{d}q}{\text{d}r}\right)^2\right]. \label{s84}
\end{eqnarray}

\twocolumngrid

Consider two light rays directed from a unique source toward an observer with coordinates $(r,p,q)$ and separated by a time-interval $\tau$, short with respect to the time travelled by the rays between the source and the observer. The trajectory of the first ray is given by
\begin{equation}
 (t,p,q) = (T(r), X(r), Y(r)).  \label{n1}
\end{equation}
The equation for the second signal is
\begin{equation}
(t,p,q) = (T(r) + \tau(r),X(r) + \zeta(r),Y(r) + \psi(r)). \label{n2}
\end{equation}
This means that, while the first ray intersects some given hypersurface $r=\text{const}$ at $(t,p,q) = (T,X,Y)$, the second ray will intersect the same hypersurface, not only later on, but, in general, at a different comoving location. Therefore, in general, the two rays will intersect different sequences of intermediate matter worldlines.

Since we are using comoving coordinates, we have $(\zeta, \psi) = (0,0)$ at the emitter and at the observer. Moreover, the directions of the first and second rays are determined by $(\text{d}p/\text{d}r,\text{d}q/\text{d}r)$ and $(\text{d}p/\text{d}r + \xi(r),\text{d}q/\text{d}r + \eta(r))$ respectively, with $\xi = \text{d}\zeta/\text{d}r$ and $\eta = \text{d}\psi/\text{d}r$ . It is assumed that $\text{d}\tau/\text{d}r, \zeta,\psi,\xi$, and $\eta$ are small, of the same order as $\tau$, hence all terms nonlinear in any of them and terms involving their products will be neglected.

Since $\zeta = \psi = 0$ at the observer, they are not observable. Anyhow, they must be monitored along the rays because they enter the equation for $\tau$ which is connected to the redshift drift $\delta z$ through
\begin{equation}
1+z+ \delta z = \frac{\text{d}(t + \tau)/\text{d}r|_{r=r_e}}{\text{d}(t + \tau)/\text{d}r|_{r=r_o}}. \label{s77}
\end{equation}
Indeed, subtracting from (\ref{s77}) the redshift defining equation written as
\begin{equation}
1+z = \frac{\text{d}t/\text{d}r|_{r=r_e}}{\text{d}t/\text{d}r|_{r=_o}}, \label{s71}
\end{equation}
and arranging, we obtain
\begin{equation}
\delta z = (1+z)\left(\frac{\frac{\text{d}\tau/\text{d}r}{\text{d}t/\text{d}r}|_{r=r_e} - \frac{\text{d}\tau/\text{d}r}{\text{d}t/\text{d}r}|_{r=r_o}}{1+\frac{\text{d}\tau/\text{d}r}{\text{d}t/\text{d}r}|_{r=r_o}}\right). \label{s79}
\end{equation}

In the following, the $\Delta$ symbol applied to some expression will denote the difference between this expression taken at $(t + \tau,r, p + \zeta,q + \psi)$ and at $(t,r,p,q)$, linearized in $(\tau,\zeta,\psi)$. By applying $\Delta$ to (\ref{s81})-(\ref{s83}), one obtains \cite{K11}
\onecolumngrid
\begin{equation}
\frac{\text{d}^2\tau}{\text{d}r^2} + \frac{\Phi_{01} \Delta \Phi_1 + \Phi_1 \Delta \Phi_{01}}{\epsilon - k} + \frac{(\Phi_{,t}^2 + \Phi \Phi_{,tt}) \Sigma \tau}{E^2} - 2 \frac{\Phi \Phi_{,t} \Delta E \Sigma}{E^3} + \frac{\Phi \Phi_{,t} \Delta \Sigma}{E^2} + \Delta U \frac{\text{d}t}{\text{d}r} + U \frac{\text{d}\tau}{\text{d}r} = 0,  \label{n4}
\end{equation}
\begin{eqnarray}
\frac{\text{d}^2\zeta}{\text{d}r^2} &+& 2\left(\frac{\Phi_{,tt}}{\Phi} - \frac{\Phi_{,t}^2}{\Phi^2}\right)\frac{\text{d}t}{\text{d}r}\frac{\text{d}p}{\text{d}r}\tau + 2\frac{\Phi_{,t}}{\Phi}\frac{\text{d}p}{\text{d}r}\frac{\text{d}\tau}{\text{d}r} + 2\frac{\Phi_{,t}}{\Phi}\frac{\text{d}t}{\text{d}r}\xi - \frac{\Delta \Phi_1 E_{12}}{(\epsilon - k)\Phi} + \frac{\Phi_{,t}\Phi_1 E_{12}\tau}{(\epsilon - k)\Phi^2} - \frac{\Phi_1 \Delta E_{12}}{(\epsilon - k)\Phi} \nonumber \\
&+& 2 \left(\frac{\Delta \Phi_1}{\Phi} - \frac{\Phi_1 \Phi_{,t} \tau}{\Phi^2}\right) \frac{\text{d}p}{\text{d}r} + 2 \frac{\Phi_1}{\Phi}\xi - \left(\frac{\zeta}{SE} - \frac{E_{,p} \Delta E}{E^2}\right) \left(\frac{\text{d}p}{\text{d}r}\right)^2 - 2\frac{E_{,p}\xi}{E}\frac{\text{d}p}{\text{d}r} - 2 \left(\frac{\psi}{SE} - \frac{E_{,q} \Delta E}{E^2}\right) \frac{\text{d}p}{\text{d}r} \frac{\text{d}q}{\text{d}r} \nonumber \\
&-& 2 \frac{E_{,q}}{E}\left(\frac{\text{d}q}{\text{d}r}\xi + \frac{\text{d}p}{\text{d}r}\eta \right) + \left(\frac{\zeta}{SE} - \frac{E_{,p} \Delta E}{E^2}\right)\left(\frac{\text{d}q}{\text{d}r}\right)^2 + 2 \frac{E_{,p}\eta}{E} \frac{\text{d}q}{\text{d}r} + \Delta U \frac{\text{d}p}{\text{d}r} + U \xi = 0, \label{n5}
\end{eqnarray}
\begin{eqnarray}
\frac{\text{d}^2\psi}{\text{d}r^2} &+& 2\left(\frac{\Phi_{,tt}}{\Phi} - \frac{\Phi_{,t}^2}{\Phi^2}\right)\frac{\text{d}t}{\text{d}r}\frac{\text{d}q}{\text{d}r}\tau + 2\frac{\Phi_{,t}}{\Phi}\frac{\text{d}q}{\text{d}r}\frac{\text{d}\tau}{\text{d}r} + 2\frac{\Phi_{,t}}{\Phi}\frac{\text{d}t}{\text{d}r}\eta - \frac{\Delta \Phi_1 E_{13}}{(\epsilon - k)\Phi} + \frac{\Phi_{,t}\Phi_1 E_{13}\tau}{(\epsilon - k)\Phi^2} - \frac{\Phi_1 \Delta E_{13}}{(\epsilon - k)\Phi} \nonumber \\
&+& 2 \left(\frac{\Delta \Phi_1}{\Phi} - \frac{\Phi_1 \Phi_{,t} \tau}{\Phi^2}\right) \frac{\text{d}q}{\text{d}r} + 2 \frac{\Phi_1}{\Phi}\eta - \left(\frac{\psi}{SE} - \frac{E_{,q} \Delta E}{E^2}\right) \left(\frac{\text{d}p}{\text{d}r}\right)^2 + 2\frac{E_{,q}\xi}{E}\frac{\text{d}p}{\text{d}r} - 2 \left(\frac{\zeta}{SE} - \frac{E_{,p} \Delta E}{E^2}\right) \frac{\text{d}p}{\text{d}r} \frac{\text{d}q}{\text{d}r} \nonumber \\
&-& 2 \frac{E_{,p}}{E}\left(\frac{\text{d}q}{\text{d}r}\xi + \frac{\text{d}p}{\text{d}r}\eta \right) - \left(\frac{\psi}{SE} - \frac{E_{,q} \Delta E}{E^2}\right)\left(\frac{\text{d}q}{\text{d}r}\right)^2 - 2 \frac{E_{,q}\eta}{E} \frac{\text{d}q}{\text{d}r} + \Delta U \frac{\text{d}q}{\text{d}r} + U \eta = 0, \label{n6}
\end{eqnarray}
where
\begin{equation}
\Phi_1 \equiv \Phi_{,r} - \Phi\frac{E_{,r}}{E},  \quad \Phi_{01} \equiv \Phi_{,tr} - \Phi_{,t}\frac{E_{,r}}{E}, \quad \Phi_{11} \equiv \Phi_{,rr} - \Phi\frac{E_{,rr}}{E}, \label{n7}
\end{equation}
\begin{equation}
E_{12} \equiv E_{,r}E_{,p} - E E_{,rp}, \quad E_{13} \equiv E_{,r}E_{,q} - E E_{,rq}, \quad \Sigma \equiv \left(\frac{\text{d}p}{\text{d}r}\right)^2 + \left(\frac{\text{d}q}{\text{d}r}\right)^2, \label{n8}
\end{equation}
\begin{equation}
 \Delta \Phi = \Phi_{,t} \tau, \quad \Delta(\Phi_{,t}) = \Phi_{,tt}\tau, \quad \Delta \frac{\text{d}t}{\text{d}r} = \frac{\text{d}\tau}{\text{d}r}, \quad \Delta \frac{\text{d}p}{\text{d}r} = \xi, \quad \Delta \frac{\text{d}q}{\text{d}r} = \eta, \label{n9}
\end{equation}
\begin{equation}
 \Delta E = E_{,p}\zeta +  E_{,q}\psi, \quad  \Delta E_{,p} = \frac{\zeta}{S}, \quad  \Delta E_{,q} = \frac{\psi}{S}, \quad \Delta \Phi_1 = \Phi_{01} \tau + \frac{\Phi E_{12}}{E^2} \zeta + \frac{\Phi E_{13}}{E^2} \psi, \label{n10}
\end{equation}
\begin{equation}
\Delta \Phi_{01} = (\Phi_{,ttr} - \Phi_{,tt}\frac{E_{,r}}{E}) \tau + \frac{\Phi_{,t} E_{12}}{E^2} \zeta + \frac{\Phi_{,t} E_{13}}{E^2} \psi, \label{n11}
\end{equation}
\begin{equation}
\Delta \Phi_{11} = (\Phi_{,ttr} - \Phi_{,t}\frac{E_{,rr}}{E}) \tau + \frac{\Phi}{E^2}\left[(E_{,rr} E_{,p} - E E_{,rrp})\zeta + (E_{,rr}E_{,q} - E E_{,rrq})\psi\right], \label{n12}
\end{equation}
\begin{equation}
\Delta E_{12} = (E_{,r} E_{,pp} - E E_{,rpp})\zeta + (E_{,rq}E_{,p} - E_{,q} E_{,rp})\psi, \quad \Delta E_{13} = (E_{,rp} E_{,q} - E_{,p} E_{,rq})\zeta + (E_{,r}E_{,qq} - E E_{,rqq})\psi,\label{n13}
\end{equation}
\begin{equation}
\Delta \Sigma = 2 \frac{\text{d}p}{\text{d}r} \xi + 2 \frac{\text{d}q}{\text{d}r} \eta,\label{n14}
\end{equation}
\begin{eqnarray}
\Delta U &=& 2 \left( - \frac{\Delta \Phi_{01}}{\Phi_1} + \frac{\Phi_{01} \Delta \Phi_1}{\Phi_1^2}\right)\frac{\text{d}t}{\text{d}r} - 2 \frac{\Phi_{01}}{\Phi_1}\frac{\text{d}\tau}{\text{d}r} - \frac{\Delta \Phi_{11}}{\Phi_1} + \frac{\Phi_{11} \Delta \Phi_1}{\Phi_1^2} + \frac{\Delta E_{,r}}{E} - \frac{E_{,r} \Delta E}{E^2} \nonumber \\
&+& 2 \left( - \frac{\Phi_{,t} E_{12} \tau}{E^2 \Phi_1} + 2 \frac{\Phi \Delta E E_{12}}{E^3 \Phi_1} - \frac{\Phi \Delta E_{12}}{E^2 \Phi_1} + \frac{\Phi E_{12} \Delta \Phi_1}{E^2 \Phi_1^2}\right) \frac{\text{d}p}{\text{d}r} - 2 \frac{\Phi E_{12} \xi}{E^2 \Phi_1} 
 \nonumber \\
&+& 2 \left( - \frac{\Phi_{,t} E_{13} \tau}{E^2 \Phi_1} + 2 \frac{\Phi \Delta E E_{13}}{E^3 \Phi_1} - \frac{\Phi \Delta E_{13}}{E^2 \Phi_1} + \frac{\Phi E_{13} \Delta \Phi_1}{E^2 \Phi_1^2}\right) \frac{\text{d}q}{\text{d}r} - 2 \frac{\Phi E_{13} \eta}{E^2 \Phi_1} \nonumber \\
&+& \frac{(\epsilon - k) \Phi \Sigma}{E^2 \Phi_1}\left(\frac{\Phi_{,t} \tau}{\Phi} - 2 \frac{\Delta E}{E} - \frac{\Delta \Phi_1}{\Phi_1} + \frac{\Delta \Sigma}{\Sigma} \right).\label{n15}
\end{eqnarray}

Then, applying the $\Delta$-operation to (\ref{s11}), written with $r$ as a parameter on the null geodesics, one obtains
\begin{equation}
\frac{\text{d}\tau}{\text{d}r} \frac{\text{d}t}{\text{d}r} = \frac{\Phi_1 \Delta \Phi_1}{\epsilon - k} + \left(\frac{\Phi \Phi_{,t}\tau}{E^2} -  \frac{\Phi^2 \Delta E}{E^3}\right) \Sigma + \frac{\Phi^2}{E^2}\left(\frac{\text{d}p}{\text{d}r}\xi  + \frac{\text{d}q}{\text{d}r} \eta \right).  \label{n16}
\end{equation}

\twocolumngrid

The redshift drift measured by the observer with coordinates $(t,r,p,q)$ in a Szekeres spacetime is determined by $\text{d}\tau/\text{d}r$ and $k^t$ through (\ref{s79}), but the null geodesic equations have to be solved together with (\ref{n4})-(\ref{n6}), since $\zeta$ and $\psi$ must be monitored along the ray because the equations which determine $(\tau, \zeta, \psi)$ are coupled and the same applies to the determination of $k^t$, depending on the evolution of $k^p$, and $k^q$. Therefore, after solving this double set of equations, we have all the results needed to obtain the measurable redshift drift from (\ref{s79}).

Moreover, the second ray is emitted in a different direction and is received from a different direction by the observer who thus sees every light source slowly drifting across the sky \cite{K11,K13}. This effect is present in other inhomogeneous cosmological models, e. g., it affects non-radial rays in spherically symmetric LTB models. It has been predicted, in this framework, under the name of cosmic parallax \cite{Q09,Q12}.

It has been demonstrated by Krasi\'nski \cite{K23} that the only spacetimes in the Szekeres family where the position drift vanishes for every null geodesic pair are the Friedmann models. Therefore, an observation of this drift for any remote source would be an evidence for the Universe to be inhomogeneous at large scales. To obtain a rough estimate of the magnitude of this effect, Krasi\'nski and Bolejko \cite{K11} have calculated the time-averaged rate of change of the source position in a LTB configuration where the null geodesics on which light travels are nonradial. They obtained a drift $\sim 10^{-6}-10^{-7}$arcsec/year. This effect is thus too tiny to be measured by current or coming soon experiments. However, since we experience regularly tremendous technical progresses, hope is left for a future probe of such a position drift.

\subsection{Cosmological blueshifts in QSS spacetimes} \label{bs}

For completeness, we stress that the possible existence of strongly blueshifted light sources in QSS spacetimes has been rather recently put forward \cite{K16}. It has been shown that, in an axially symmetric QSS model, infinite blueshift can appear, but only on axial rays, i. e., rays which intersect every space orthogonal to the dust flow on the symmetry axis. Moreover, using a particular axially symmetric QSS model, strong blueshifts for rays emitted shortly after the Big Bang and running close to the symmetry axis have been numerically obtained. Then, in a QSS toy model deprived of any symmetry, two null lines such that rays in their vicinity exhibit similar features to those in the vicinity of axial null geodesics in axially symmetric QSS models have been numerically found. This hints at the tentative possibility that rays observed with infinite or very strong blueshifts exist in general QSS spacetimes and are focussed around two special directions depending on the particular features of the model.

However, the Szekeres model validity region is bounded by the matter-radiation equality surface, since its gravitational fluid source, being dust, cannot account for any pressure component. Therefore, to move the initial frequencies of the relic radiation to an observable range, a particular form must be imposed to the $t_B(r)$ Bang function in a QSS region with no origin \cite{K20}. Hence, a possible future observation of such an effect would set additional stringent constraints on the Szekeres model to be fitted by other data, i. e., the quasispherical nature of the region of interest, implying there $(\epsilon = +1)$, and the location and amplitude of the hump prescribed for the $t_B(r)$ function.

\section{Processing the data fitting with deep learning} \label{dl}

The Szekeres equations displayed in this paper cannot be solved analytically. In previous studies involving these spacetimes, the method used was: assuming a set of given expressions for the Szekeres functions and parameters written with a finite number of constant parameters, fit these constant parameters to some cosmological data of interest and claim that the matched model solves the issue encompassed by the data \cite{I08,B09,BC10,B11,B13,T13,I13,P14,B16,M22}. In these works, the Szekeres equations are numerically integrated.

Now, with our more general model, besides one constant parameter, $\Lambda$, and one quasiconstant parameter, $\epsilon$, we have also five functions of the $r$ coordinate which are to be determined by the data. However, some simplifying constraints have to be imposed on these determining constants and functions. They are described below.

\subsection{Basic physical constraints}

To preserve the $(-+++)$ metric signature, we must have
\begin{equation}
\epsilon - k > 0. \label{dl1}
\end{equation}
This implies that, for quasihyperbolic regions $(\epsilon = -1)$, only the hyperbolic evolution, obtained for $k<0$, is possible, while quasiplanar regions can only experiment parabolic or hyperbolic evolution, where $k \leq 0$, and quasispherical regions can display the three types of evolution, elliptic ($k>0$), parabolic ($k=0$), and hyperbolic ($k<0$). Such a crossed-constraint will have to be checked {\it a posteriori} for each cosmological model issuing from any data fit.

Shell-crossings are loci in spacetimes where the energy density diverges. They must therefore be avoided in a well-behaved cosmological model. The conditions for avoiding shell-crossing surfaces in quasi-spherical regions have been worked out by Szekeres \cite{S75b} and enlarged to the other two classes of models by Hellaby and Krasi\'nski \cite{H08,H02}. These requirements must be fulfilled by any quasispherical region in any proper Szekeres cosmological model and this will have to be checked also {\it a posteriori}.

For obvious physical reasons, the areal radius $\Phi$ has to be positive, which reads
\begin{equation}
\Phi(t,r) > 0, \quad \forall t \quad \text{and} \quad r. \label{dl2}
\end{equation}

The metric should be nondegenerate and nonsingular, except at the bang time $t_B(r)$. Since $(\text{d}p^2 + \text{d}q^2)/E^2$ maps to the unit sphere, plane, or pseudosphere, , $|S(r)| \neq 0$ is needed for a sensible mapping, and therefore, we can choose $S>0$. In the cases where $\epsilon = 0$ or $1$, $E$ is bound to vanish at some $(p,q)$ location. Thus the mapping is badly behaved there and the presence of such loci in a Szekeres region would rule out the corresponding model.

A well-behaved $r$ coordinate implies
\begin{equation}
\infty > \frac{(\Phi_{,r} - \Phi E_{,r}/E)^2}{\epsilon -k} >0. \label{dl3}
\end{equation}

In accordance with the weak energy condition, the energy density must be positive and the Kretschmann scalar must be finite which yields
\begin{equation}
\text{either} \quad M_{,r} - 3 M E_{,r}/E \geq 0 \quad \text{and} \quad \Phi_{,r} - \Phi E_{,r}/E \geq 0, \label{dl4}
\end{equation}
\begin{equation}
\text{or} \quad M_{,r} - 3 M E_{,r}/E \leq 0 \quad \text{and} \quad \Phi_{,r} - \Phi E_{,r}/E \leq 0. \label{dl5}
\end{equation}

Finally, the asymptotic recovery of an FLRW universe at the scales of inhomogeneity/homogeneity transition sets constraints on the forms of the functions to be fitted. Indeed, the Friedmann limit follows \cite{K12} when $\Phi(t,r) = r R(t)$, $k = k_0 r^2$, with $k_0 = \text{const}$, $B(r) = C(r) = 0$, and $D(r) = 4A(r) = \epsilon$,
where
\begin{equation}
E(r,p,q) = A(p^2 + q^2) + 2 Bp + 2 Cq + D. \label{dl6}   
\end{equation}
Using the $E$ definition (\ref{s2}), this gives, for the most general Szekeres model where $\epsilon$ can be any among $\{-1,0,+1\}$,
\begin{equation}
S = 2 \epsilon, \quad P=Q=0, \quad  \label{dl8}   
\end{equation}
and still
\begin{equation}
\Phi(t,r) = r R(t), \quad k = k_0 r^2.
\label{dl9}   
\end{equation}

Therefore, we have five constraints to be fulfilled by any Szekeres cosmological model, which are summarized in (\ref{dl8}) and (\ref{dl9}) where $t\geq t_{trans}$ and $r \geq r_{trans}$ with the subscript 'trans' implying that the quantity is evaluated at the inhomogeneity/homogeneity transition. We are thus provided with two extra constant parameters for our model, $t_{trans}$ and $r_{trans}$, which will have either to be determined through data fitting or, as a temporary simplification, fixed by an {\it a priori} but physically justified choice.

Now, although the choice of the way to determine the $r$ coordinate is wide, it might be comfortable, for practical purpose, to choose a definition which includes readily the Friedmann limit, e. g., by setting $k(r) = k_0 r^2$ everywhere. In this case, only $k_0$ is left as a constant parameter to be determined by the observations. However, other choices can be convenient. For example, $t_B(r) = t_0$ implies that the bang time is constant in the whole spacetime, which complies with Friedmann. One can, without loss of generality, equalize $t_0$ to zero, which amounts to a mere rescaling of the time coordinate.  However, should the blueshift effect discussed in Sec.\ref{bs} be observed in the future, a more complicated bang function would arise. In the FLRW limit we have also, from (\ref{s3}), $E/r^2$ and $M/r^3$ constant \cite{K12}. Since, in the generic Szekeres solution, $E$ is a function of $r$ but also of $p$ and $q$, a choice of the type $E= E_0 r^2$ might be too restrictive and might suppress some of the most interesting properties of the Szekeres solutions, while $M = M_0 r^3$ can be a better choice.

\subsection{Fitting method}

The problem is now to determine a cosmological model depending on five functions of $r$, chosen among $S, P, Q, M$, $k$, and $t_B$, and on seven constant parameters: $\epsilon$, $\Lambda$, $t_{trans}$, $r_{trans}$, and the spatial coordinates of the observer, $r_{obs}$, $p_{obs}$, and $q_{obs}$.

For this, we have, in principle, an enormous set of data from surveys which are designed each to probe one or more particular cosmological effect. To deal with these data, we need a device very powerful and a method adapted to the fitting of functions. Both now exist. The device is one of those machines designed to implement deep learning and the method is symbolic regression.

Symbolic regression aims at determining among a huge set of data points the best-fit curve representing the data behavior. The results are obtained using machine-learning and they can be displayed either numerically or analytically depending on the problem at hand. This method can be employed to fit the Szekeres functions through a comparison of the model predictions to the actual cosmological data.

The standard way of completing the training of machine learning tools is to have them running on already solved problems. Some are available in the literature, others will have to be produced on purpose. Such solutions currently at hand have usually been obtained by numerically reproducing a given set of measured cosmological data through simplified forms of the Szekeres functions.

As an interesting example, which might be used among a number of others for neural network training, we are going to analyze the model employed in \cite{B16}. There, the authors simulate the local expansion of the Universe with Szekeres solutions which match the standard FLRW model on $\geq 100 h^{-1}$ Mpc scales but exhibit nonkinematic relativistic differential expansion on small scales, while being consistent with Planck CMB temperature anisotropies. They write therefore the Szekeres mass functions as exhibiting a small departure from the FLRW limit and including features designed to account for both the Local Void and the Great Attractor. They set the age of the Universe to be everywhere the same for a comoving observer and equal to that of an asymptotic background spatially flat FLRW model. Then, they assume axial symmetry, with a Szekeres dipole described by the function $S$ only. Thus, the model has seven free parameters: four parameters that specify the Szekeres model and three parameters that specify the position of the observer.
Finally the authors restrict their models to five parameters and restrain it to be consistent with the observed CMB temperature anisotropies, while simultaneously fitting the five parameters to the redshift variation of the Hubble expansion dipole.

The above example shows how physically motivated training models can be constructed. However, let us discuss the different assumptions made.

Axial symmetry is indeed an interesting mathematical simplification, since spacetimes exhibiting this symmetry, possess repeatable light paths, in the sense of \cite{K11}, which are the null geodesics intersecting every space of constant time on the axis of symmetry. However, models with such a symmetry are not well suited to describe our universe whose matter distribution does not seem to appear axisymmetric. Moreover, in such spacetimes, the mass dipole should be measured in only one direction, which could be a final test for axial symmetry. Therefore, even though this property might be chosen for a number of trainings, their set should be completed by non-axially symmetric spacetimes.

The vanishing of the $t_B(r)$ function is, on the contrary, a valid assumption, since the initial singularity being outside the matter dominated region, at that locus, the FLRW limit applies.

The consideration of a $\delta M(r)$ perturbation is also relevant. It plays the role of the density contrast in standard cosmology, even though the usual tools of perturbation theory are not appropriate here, see Sec.\ref{nc}.

The vanishing of the two dipole functions $P$ and $Q$ is related to the axial symmetry of the model. Indeed, $S$, $P$, and $Q$ are solely responsible for the anisotropy, each defining it in an orthogonal direction. A non-vanishing $S$ only, therefore determines the anisotropy in the axial direction and implies the symmetry. For non-axially symmetric training models, this feature will have to be dropped.

A number of mock catalogues of the different data sets analyzed in the present paper will thus have to be generated for toy Szekeres models exhibiting different functions with different values for their free defining parameters. For training purpose, it is not mandatory that these catalogues reproduce actual observed data. It suffices that they should be obtained from given Szekeres models of different kinds.

The ultimate goal will then be to use the whole set of available cosmological data to thoroughly constrain the model. The aim of the present paper is therefore to give the most comprehensive set of tools needed to complete such a task in a Szekeres framework. However, we are aware that this will take a rather long time being carried out. But deep learning begins only to make a few steps in cosmology \cite{L24} and it should be appropriate that these steps should be done in the right direction.

We want also to stress that it is a current standard practice to automatically transform redshift data to the CMB rest frame before performing any analysis. The canonical CMB rest frame is defined by matching the 3.37 mK temperature of the CMB dipole to the dipole in a series expansion adapted to the $\Lambda$CDM paradigm. For a fully model independent study, this practice should be given up and the data used for our purpose should be the rawest as possible. This implies also to remove any $\Lambda$CDM inspired step from the data processing, such as in bias or foreground removals.

\section{Conclusions} \label{concl}

We have provided, in this paper, a thorough analysis of the Szekeres solution as a cosmological model. This solution benefits indeed from being devoid of any symmetry and, as an exact GR solution for dust plus a cosmological constant, it can be used as a substitute to the perturbed $\Lambda$CDM homogeneous model to describe the late Universe. At larger scales, it can be joined smoothly to a Friedmannian early time Universe which it has the nice property to include as a limit. We have here summarized, updated, corrected when needed, and completed with new results thirty years of work on the cosmological abilities of the Szekeres solutions. Actually, our aim has been to provide generalized, comprehensive, and consistent tools which can be readily used for future broad cosmological studies conducted with a well-adapted GR solution.

We have therefore presented the most general form of the equations needed to calculate the cosmological quantities which will have to be confronted to the different data surveys. We have displayed the light ray trajectories or null geodesics, the redshift, the different cosmological distances (area, luminosity, etc.), the intrinsic matter dipole, the BAO scale, the weak-lensing parameters (convergence and shear).Then we have proposed methods to use the current or soon-coming data surveys as probes to constraint the Szekeres independent parameters and functions. The effects which have been analyzed here are supernova dimming, galaxy number count evolution, CMB multipoles, SZ effects, BAO, weak lensing, redshift space distortions, redshift drift, differential cosmic expansion, position drift, and possible blueshift effect. In particular, the cosmographic formalism has been thoroughly adapted to the Szekeres framework and, from the expressions obtained for the multipole components of the series expansion pertaining to this formalism, we have been able to show that the Szekeres model can reproduce the main multipole features as seen in different supernova observations. This constitutes a new key result in favor of the Szekeres formalism. The ultimate goal of finding the Szekeres model representing accurately the late Universe would imply that, at least, seven probes testing different physical effects should be used, one for each degree of freedom of the Szekeres parameters and functions. Given the heap of in principle available data, we have suggested to use machine learning to tackle more efficiently this issue.

The program described in this paper could appear rather ambitious. However, it seems to be one of the best ways to deal with the cosmological problem and to get rid of the current tensions and anomalies which plague the FLRW paradigm, as far as one wishes to keep GR as the proper gravitation  theory, which is the main underlying physical paradigm of this endeavour. Moreover, the results displayed here can be of valuable use for any stepwise undertaking.

\acknowledgements

The author wishes to acknowledge enlightening discussions with Jean-Michel Alimi and with Andrzej Krasi\'nski and very constructive comments and suggestions from the anonymous referee.

\onecolumngrid

\appendix
\section{Szekeres geometrical quantities of interest}

For completeness and to correct some errors which can be found in the literature \cite{T14}, we provide here the expressions for the Christoffel symbols, the components of the Ricci and of the Weyl tensors pertaining to the Szekeres solutions and needed to complete the calculations described in this paper. These quantities have been obtained by running the SageMath software \cite{SM} with the simplifying notations of (\ref{s53}).

The nonzero Christoffel symbols are
\begin{equation}
\Gamma^t_{rr} = H H_{,t} \quad \Gamma^t_{pp} = \Gamma^t_{qq} = F F_{,t}, \quad \Gamma^r_{rt} = \frac{H_{,t}}{H}, \quad \Gamma^r_{rr} = \frac{H_{,r}}{H}, \quad \Gamma^r_{rp} =\frac{H_{,p}}{H}, \quad \Gamma^r_{rq} = \frac{H_{,q}}{H},\nonumber
\end{equation}
\begin{equation}
\Gamma^r_{pp} = \Gamma^r_{qq} = - \frac{F F_{,r}}{H^2}, \quad \Gamma^p_{pt} = \Gamma^q_{qt} = \frac{F_{,t}}{F}, \quad \Gamma^p_{rp} = \Gamma^q_{rq} = \frac{F_{,r}}{F}, \quad \Gamma^p_{rr} = - \frac{H H_{,p}}{F^2}, \quad \Gamma^q_{rr} = - \frac{H H_{,q}}{F^2}, \nonumber
\end{equation}
\begin{equation}
\Gamma^p_{pq} = \Gamma^q_{qq} = - \Gamma^q_{pp} = \frac{F_{,q}}{F}, \quad \Gamma^p_{pp} = - \Gamma^p_{qq} = \Gamma^q_{pq} = \frac{F_{,p}}{F}. \nonumber
\end{equation}

The nonzero components of the Ricci tensor are
\begin{equation}
R_{tt} = - \left(\frac{H_{,tt}}{H} +2 \frac{F_{,tt}}{F}\right), \quad R_{tr} = 2 \left(\frac{H_{,t} F_{,r}}{H F} - \frac{F_{,tr}}{F}\right), \nonumber
\end{equation}
\begin{equation}
R_{tp} = - \frac{F_{,tp}}{F} - \frac{H_{,tp}}{H} + \frac{F_{,t} F_{,p}}{F^2} + \frac{H_{,p} F_{,t}}{H F}, \quad R_{tq} = - \frac{F_{,tq}}{F} - \frac{H_{,tq}}{H} + \frac{F_{,t} F_{,q}}{F^2} + \frac{H_{,q} F_{,t}}{H F}, \nonumber
\end{equation}
\begin{equation}
R_{rr} = 2 \frac{H H_{,t} F_{,t}}{F} + H H_{,tt} - 2 \frac{F_{,rr}}{F} + 2 \frac{H_{,r} F_{,r}}{H F} - \frac{H}{F^2}\left(H_{,pp} + H_{,qq}\right), \nonumber
\end{equation}
\begin{equation}
R_{rp} = - \frac{F_{,rp}}{F} + \frac{F_{,r} F_{,p}}{F^2} + \frac{H_{,p} F_{,r}}{H F}, \quad R_{rq} = - \frac{F_{,rq}}{F}  + \frac{F_{,r} F_{,q}}{F^2} + \frac{H_{,q} F_{,r}}{H F}, \nonumber
\end{equation}
\begin{eqnarray}
R_{pp} &=& (F_{,t})^2 + F F_{,tt} + \frac{F H_{,t} F_{,t}}{H} - \frac{(F_{,r})^2}{H^2} - \frac{F F_{,rr}}{H^2} + \frac{(F_{,p})^2}{F^2} \nonumber \\
&-& \frac{F_{,pp}}{F} + \frac{(F_{,q})^2}{F^2} - \frac{F_{,qq}}{F} + \frac{F H_{,r} F_{,r}}{H^3} + \frac{H_{,p} F_{,p}}{H F} - \frac{H_{,pp}}{H} - \frac{H_{,q} F_{,q}}{H F}, \nonumber
\end{eqnarray}
\begin{equation}
R_{pq} = \frac{H_{,p} F_{,q}}{H F} - \frac{H_{,pq}}{H} + \frac{H_{,q} F_{,p}}{H F}, \nonumber
\end{equation}
\begin{eqnarray}
R_{qq} &=& (F_{,t})^2 + F F_{,tt} + \frac{F H_{,t} F_{,t}}{H} - \frac{(F_{,r})^2}{H^2} - \frac{F F_{,rr}}{H^2} + \frac{(F_{,p})^2}{F^2} \nonumber \\
&-& \frac{F_{,pp}}{F} + \frac{(F_{,q})^2}{F^2} - \frac{F_{,qq}}{F} + \frac{F H_{,r} F_{,r}}{H^3} - \frac{H_{,p} F_{,p}}{H F} - \frac{H_{,qq}}{H} + \frac{H_{,q} F_{,q}}{H F}. \nonumber
\end{eqnarray}

The non-zero components of the Weyl tensor are
\begin{eqnarray}
C_{trtr} &=& \frac{H}{3 F^2} \left\{ -H (F_{,t})^2 + H F F_{,tt} + F H_{,t} F_{,t} - F^2 H_{,tt} +  \frac{(F_{,r})^2}{H} - \frac{F}{H} F_{,rr} \right. \nonumber \\
&-& \left. \frac{H}{F^2}\left[(F_{,p})^2 + (F_{,q})^2\right] + \frac{H}{F}(F_{,pp} + F_{,qq}) + \frac{F H_{,r} F_{,r}}{H^2} - \frac{1}{2}(H_{,pp} + H_{,qq}) \right\}, \nonumber
\end{eqnarray}
\begin{equation}
C_{trtp} = \frac{1}{2 F} \left( - F_{,rp} + \frac{F_{,r} F_{,p}}{F} + \frac{H_{,p} F_{,r}}{H}\right), \quad C_{trtq} = \frac{1}{2 F} \left( - F_{,rq} + \frac{F_{,r} F_{,q}}{F} + \frac{H_{,q} F_{,r}}{H}\right), \nonumber
\end{equation}
\begin{eqnarray}
C_{tptp} &=& \frac{1}{6} \left\{ (F_{,t})^2 - F F_{,tt} - \frac{F H_{,t} F_{,t}}{H} + \frac{F^2 H_{,tt}}{H} -  \frac{(F_{,r})^2}{H^2} + \frac{F F_{,rr}}{H^2} + \frac{1}{F^2}\left[(F_{,p})^2 + (F_{,q})^2\right] \right. \nonumber \\
&-& \left. \frac{1}{F}(F_{,pp} + F_{,qq}) - \frac{F H_{,r} F_{,r}}{H^3} + \frac{3}{H F}(H_{,p} F_{,p}- H_{,q} F_{,q}) - \frac{H_{,pp}}{H} + 2 \frac{H_{,qq}}{H}\right\}, \nonumber
\end{eqnarray}
\begin{equation}
C_{tptq} = \frac{1}{2 H} \left( - H_{,pq} + \frac{H_{,p} F_{,q}}{F} + \frac{H_{,q} F_{,p}}{F}\right), \nonumber
\end{equation}
\begin{eqnarray}
C_{tqtq} &=& \frac{1}{6} \left\{ (F_{,t})^2 - F F_{,tt} - \frac{F H_{,t} F_{,t}}{H} + \frac{F^2 H_{,tt}}{H} -  \frac{(F_{,r})^2}{H^2} + \frac{F F_{,rr}}{H^2} + \frac{1}{F^2}\left[(F_{,p})^2 + (F_{,q})^2\right] \right. \nonumber \\
&-& \left. \frac{1}{F}(F_{,pp} + F_{,qq}) - \frac{F H_{,r} F_{,r}}{H^3} + \frac{3}{H F}(- H_{,p} F_{,p}+  H_{,q} F_{,q}) + 2 \frac{H_{,pp}}{H} - \frac{H_{,qq}}{H}\right\}, \nonumber
\end{eqnarray}
\begin{eqnarray}
C_{rttr} &=& \frac{H^2}{3 F^2} \left\{ (F_{,t})^2 - F F_{,tt} - \frac{F H_{,t} F_{,t}}{H} + \frac{F^2 H_{,tt}}{H} -  \frac{(F_{,r})^2}{H^2} + \frac{F F_{,rr}}{H^2} + \frac{1}{F^2}\left[(F_{,p})^2 + (F_{,q})^2\right] \right. \nonumber \\
&-& \left. \frac{1}{F}(F_{,pp} + F_{,qq}) - \frac{F H_{,r} F_{,r}}{H^3} + \frac{1}{2 H} (H_{,pp} +H_{,qq})\right\}, \nonumber
\end{eqnarray}
\begin{equation}
C_{trpr} = \frac{H^2}{2 F} \left(F_{,tp} - \frac{F_{,t} F_{,p}}{F} - \frac{F H_{,tp}}{H} + \frac{H_{,p} F_{,t}}{H}\right), \quad C_{trqr} = \frac{H^2}{2 F} \left(F_{,tq} - \frac{F_{,t} F_{,q}}{F} - \frac{F H_{,tq}}{H} +\frac{H_{,q} F_{,t}}{H}\right), \nonumber
\end{equation}
\begin{equation}
C_{rppr} = H^2 C_{tptp} -\frac{H}{F}\left(H_{,p} F_{,p} - H_{,q} F_{,q}\right) + \frac{H}{2}(H_{,pp} - H_{,qq}), \nonumber
\end{equation}
\begin{equation}
C_{rqqr} = H^2 C_{tqtq} +\frac{H}{F}\left(H_{,p} F_{,p} - H_{,q} F_{,q}\right) - \frac{H}{2}(H_{,pp} - H_{,qq}), \nonumber
\end{equation}
\begin{equation}
C_{rpqr} = - F^2 C_{tptq}, \quad C_{rpqp} = F^2 C_{trtq}, \quad C_{rqpq} = F^2 C_{trtp}, \quad C_{qpqp} = - \frac{F^4}{H^2} C_{trtr}. \nonumber
\end{equation}

\twocolumngrid

\end{document}